\documentclass[12pt]{article}
\usepackage{latexsym}
\usepackage{epsfig,graphics}

\newcommand{\be}{\begin{equation}}
\newcommand{\ee}{\end{equation}}

\def\lsi{\raise0.3ex\hbox{$<$\kern-0.75em\raise-1.1ex\hbox{$\sim$}}}
\def\gsi{\raise0.3ex\hbox{$>$\kern-0.75em\raise-1.1ex\hbox{$\sim$}}}

\begin{document}

\begin{titlepage}

\null\vspace{-3.5cm}

\begin{tabbing}
\` Oxford OUTP-01-32P \\
\end{tabbing}

\begin{centering} 

\null\vspace{1.0cm}

{\large \bf  Confining strings in SU(N) gauge theories}

\vspace{1.25cm}
B. Lucini and M. Teper

\vspace{0.8cm}
{\it Theoretical Physics, University of Oxford, 1 Keble Road, \\
Oxford OX1 3NP, UK\\}

\vspace{1.5cm}
{\bf Abstract}
\end{centering}

\noindent
We calculate the string tensions of $k$-strings in 
SU($N$) gauge theories
in both 3 and 4 dimensions. We do so for SU(4) and SU(5) in D=3+1, 
and for SU(4) and SU(6) in D=2+1. In D=3+1, we find that the ratio 
of the $k=2$ string tension to the $k = 1$ fundamental string 
tension is consistent, at the $2\sigma$ level, with both
the M(-theory)QCD-inspired conjecture that 
$\sigma_{k} \propto \sin(\pi k/N)$ and with `Casimir scaling', 
$\sigma_{k} \propto k (N-k)$. In D=2+1, where our results are
very precise, we see a definite deviation from the MQCD formula, 
as well as a much smaller but still significant deviation 
from Casimir scaling. We find that in both D=2+1 and D=3+1 the 
high temperature spatial $k$-string tensions also satisfy approximate 
Casimir scaling. We point out that approximate Casimir scaling 
arises naturally if the cross-section of the flux tube
is nearly independent of the flux carried, and that this 
will occur in an effective dual superconducting description, 
if we are in the deep-London limit. We estimate, numerically, the 
intrinsic width of $k$-strings in D=2+1 and indeed find little 
variation with $k$. In addition to the stable $k$-strings 
we investigate some of the unstable strings, which
show up as resonant states in the string mass spectrum. While in D=3+1
our results are not accurate enough to extract the string tensions of
unstable strings, our more precise calculations in D=2+1 show that there
the ratios between the tensions of unstable strings and the tension of the
fundamental string are in reasonably good agreement with (approximate) 
Casimir scaling. 
We also investigate the basic assumption that confining flux tubes
are described by an effective string theory at large distances,
and attempt to determine the corresponding universality class.
We estimate the coefficient of the universal L\"uscher correction
from periodic strings that are longer than 1 fermi, and find
$c_L=0.98(4)$ in the D=3+1 SU(2) gauge theory and $c_L=0.558(19)$ in 
D=2+1. These values are within $2\sigma$ of the simple 
bosonic string values,
$c_L=\pi/3$ and $c_L=\pi/6$ respectively, and are inconsistent 
with other simple effective string theories such as 
fermionic, supersymmetric or Neveu-Schwartz.

\vspace{0.5cm}

\vfill

\end{titlepage}

\section{Introduction}
\label{sec_intro}

It is widely believed that the SU(3) gauge theory that 
underlies QCD is linearly confining and that this explains 
why we do not observe quarks (or gluons) in nature.
The fact that confinement is linear suggests that
the colour-electric flux between fundamental charges
is localised in a tube between those charges and it
is attractive to think that the long-distance physics of
such flux tubes is given by an effective string theory.
The simplest possibility is that this string theory is 
bosonic but other possibilities are not excluded and
indeed might  be natural if QCD is obtained by some kind
of reduction from a higher-dimensional theory.
 
The same comments apply to SU($N$) gauge theories for $N\not=3$.
Indeed there are long-standing ideas that for $N\to\infty$
the SU($N$) gauge theory can be thought of as a string theory.
Moreover SU($N$) gauge theories in D=2+1 also appear to be linearly
confining 
\cite{mtd3}
and all the above comments will apply there as well.

In addition to charges in the fundamental representation 
(like quarks) one can consider the potential between static 
charges in higher representations of the gauge group. In SU(2)
and SU(3) any such charge can be screened by gluons
either to the fundamental or to the trivial representation. 
Since virtual gluons are always present in the vacuum
this means that such a potential will, at large distances, 
either rise linearly with a string tension equal to the
fundamental one or will flatten off to some constant value.
(This assumes that the fundamental string tension, $\sigma$,
is the lowest,
as appears to be the case.) For $N\geq 4$, however, this is
no longer the case and there are new stable strings with
string tensions different from the fundamental one.
A typical source may be thought of as $k$ fundamental
charges located at a point. The confining string is then
usually referred to as a $k$-string. For SU($N$) we have
non-trivial stable $k$-strings up to a maximum value of $k$
given by the integer part of $N/2$.

Such $k$-strings are interesting for a variety of reasons. The 
values of their string tensions, $\sigma_k$, will constrain models
of confinement. In models of glueballs in which the latter
consist of open or closed strings, the SU($N$) mass spectrum
should change with $N$ in a way that is determined by
how $\sigma_k$ varies with $N$ and $k$. In addition there
are theoretical ideas concerning the value of $\sigma_k$.
In particular there is a conjecture based on M-theory approaches 
to QCD (MQCD) 
\cite{mqcd}
that suggests $\sigma_k \propto \sin\{\pi k/N\}$. 
One can contrast this with the old ``Casimir scaling'' conjecture
\cite{oldcs}
that would suggest $\sigma_k \propto  k(N-k)$ and also with the simple
possibility that a $k$-string consists of $k$ non-interacting
fundamental strings, in which case $\sigma_k = k \sigma$.

In a string theory the mass of a flux tube of length $l$
will receive a leading large-$l$ correction that is $O(1/l)$.
Such a slowly decreasing correction cannot be made negligible simply 
by making $l \gg 1 \mathrm{fm}$ and so it will, in principle, limit the
accuracy of our calculations of $\sigma_k$. 
Fortunately this leading string correction is known to be universal
\cite{Luscherstring}
in that its coefficient is determined entirely by the central charge
of the effective string theory. The universality class is 
usually thought to be that of a simple Nambu-Goto bosonic string.
There is however no strong direct (numerical) evidence for this
belief that we are aware of. Such evidence would need to be obtained
from strings that are longer than $1 \ \mathrm{fm}$ and to achieve the
required accuracy for such strings is a hard numerical problem.
Where accurate values are quoted in the literature they typically 
involve
fitting potentials down to shorter distances, where the fits are
almost certainly dominated by the tail of the Coulomb term which
has the same functional form as the string correction (and in
practice a similar coefficient).
We have therefore attempted to provide a usefully accurate 
calculation of this string correction in SU(2) gauge theory, 
in both D=3+1 and D=2+1.
Such a calculation also addresses the fundamental question
of whether a confining flux tube is in fact described by an
effective string theory at large distances.

The contents of this paper are as follows. In the next
Section we describe how we calculate $\sigma_k$ from the mass 
of a flux loop that winds around the spatial torus. We contrast
this method with one that uses explicit sources; in particular 
for `strings' that can break. All this requires
a classification of strings in all possible representations
of the gauge group (the details of which appear in
Section~{\ref{a1} of the Appendix). 
In the following Section we summarise the lattice aspects of our 
calculation; we can be brief since it is entirely standard. 
We then turn to the basic question of whether we really do have strings
and, if so, which universality class they belong to. Confining
ourselves to flux tubes that are longer 
than $1 \mathrm{fm}$ we find that
in both  D=3+1 and D=2+1 SU(2) gauge theory the leading correction 
to the linear dependence of the string mass is consistent, within
quite small errors, with what one expects from the simplest 
effective bosonic string theory and excludes the most obvious 
alternatives. We then turn to our calculation of $k$-strings.
We begin by briefly summarising some of the theoretical expectations:
MQCD, Euclidean and Hamiltonian (lattice) strong
coupling, Casimir scaling, the bag model and simple 
flux counting. We then turn to our D=3+1 calculations of 
$\sigma_{k=2}$ in both SU(4) and SU(5) gauge theories
and follow this with our (inevitably) much more accurate D=2+1 
calculations for SU(4) and SU(6). (In SU(6) we are able 
to address non-trivial $k=3$ strings.) In D=3+1 we find consistency
at the $2\sigma$ level
with both MQCD and Casimir scaling. In D=2+1 the string tension 
ratios, while still close to the MQCD formula, are much 
closer to Casimir scaling. 
We point out that if the flux is homogeneous, then
(approximate) Casimir scaling arises if the flux tube width is 
(approximately) independent of $k$. (And, more theoretically,
that this will arise in the deep-London limit of a dual
superconducting vacuum.) To test this 
idea we perform an explicit calculation of the intrinsic
size of $k$-strings in D=2+1. We find that 
the $k$-string width is indeed largely independent of $k$,
albeit with some interesting if weak differences.
We then point out that the
same calculations can be reinterpreted as telling us that the
spatial string tension in the high temperature deconfining
phase satisfies approximate Casimir scaling. We complement this
with an explicit D=3+1 high-$T$ calculation that demonstrates that
in that case too the string tension ratio is close to
Casimir scaling. We then attempt to see if there is any
sign of other, unstable, strings, which should appear as excited states
in the string mass spectrum. We find reasonably convincing evidence for such
strings, satisfying approximate Casimir scaling, in our
D=2+1 calculations. We finish with a discussion of our results
and some of their implications.

A preliminary version of our D=3+1 calculations has appeared in 
\cite{string00}.
In the introduction to a recent companion paper on the mass spectrum 
and topological properties of D=3+1 SU($N$) gauge theories 
\cite{glue01}
we briefly summarised some of our results on $\sigma_{k}/\sigma$.
In particular we drew attention to the relevance 
of these results on $k$-strings for the Casimir scaling hypothesis. 
We remark that all these calculations are intended as a first step 
to a much more complete and accurate calculation of the properties of
SU($N$) gauge theories for all values of $N$.

\section{Strings and string breaking}
\label{sec_strings}

Consider a static source in some representation $\cal R$ 
of the gauge group in, say, 3+1 dimensions. 
Suppose we have a conjugate source
a distance $r$ away. If $r$ is small then the potential
energy will be dominated by the Coulomb term
\be
V_{\cal R}(r) \stackrel{r\to 0}{=} 
{{C_{\cal R} \alpha_s(r)} \over {r}} + \ldots 
\label{eqn_potcoul}
\ee
where $\alpha_s(r)$ is the usual running coupling and
$C_{\cal R}$ is the quadratic Casimir of the representation
${\cal R}$:
\be
C_{\cal R} \equiv Tr_{\cal R} T^a T^a
\label{eqn_qcasimir}
\ee
with the $T^a$ being the generators of the group. If the 
theory is linearly confining, and if we ignore the fact 
that the source may be screened by gluons, then at large $r$ 
we expect the potential energy to be given by
\be
V_{\cal R}(r) \stackrel{r\to\infty}{=} 
\sigma_{\cal R} r - 
{{\pi(D-2)}\over{24}}{c_s\over r} + \ldots .
\label{eqn_pot}
\ee
Here  $\sigma_{\cal R}$ is the `string tension' of the
confining flux tube joining the sources and how its value
varies with the representation ${\cal R}$ is an 
interesting physical question. If the long-distance
physics of the confining flux tube is described by an
effective string theory, then the $O(1/r)$ correction in
eqn(\ref{eqn_pot}) is the Casimir energy of a string
with fixed ends, and $c_s$ is proportional to the central 
charge. This correction is universal 
\cite{Luscherstring},
since it depends only upon the massless modes in the effective 
string theory and does not depend upon the detailed
and complicated dynamics of the flux tube on scales comparable 
to its width. The central charge is given 
\cite{Polchinski}
by the number of massless bosonic and fermionic modes 
that propagate along the string. In practice it is
usually assumed that $c_s=1$, corresponding to the simplest 
possible (Nambu-Goto) bosonic string theory. However, these
modes are not related to the fundamental degrees of freedom
of our SU($N$) gauge theory in any transparent way and the presence
of fermionic modes is certainly not excluded. For example,
we have the following simple possibilities
\cite{csvarious}:
\be
c_s 
=
\left\{ \begin{array}{ll}
1   & \ \ \ {\mathrm{bosonic}} \\
\frac{1}{4}  & \ \ \ {\mathrm{fermionic}} \\
0     & \ \ \ {\mathrm{supersymmetric}} \\
\frac{3}{2}   & \ \ \ {\mathrm{Neveu-Schwartz}}
\end{array}
\right. 
\label{eqn_csvarious}
\ee
Whether a string description of the confining flux tube
is in fact valid and, if so, what is its universality class,
are fundamental questions which are still largely open.
The examples in eqn(\ref{eqn_csvarious}) show that one needs to 
calculate $c_s$ to better than, say, $\pm 15\%$ if one is to
usefully resolve different possibilities.

In reality the vacuum contains virtual gluons which can
screen the static source, and this will complicate any attempt
to calculate $\sigma_{\cal R}$. When and how this happens will
depend on the energetics of the system. Suppose the
representation  $\cal R$ can be screened to a different
representation  ${\cal R}^{\prime}$ by a number of gluons. 
Such a screened source will acquire an extra mass of, say, $\Delta M$.
If the string tension corresponding to  ${\cal R}^{\prime}$
is smaller than  $\sigma_{\cal R}$, the screening is certain
to become energetically favoured for sufficiently large $r$,
since as $r\to\infty$
\be
\Delta M \ll 
V_{\cal R}(r) - V_{{\cal R}^{\prime}}(r)
{\simeq}
\sigma_{\cal R}r - \sigma_{{\cal R}^{\prime}}r .
\label{eqn_break}
\ee
The minimum value of $r$ at which the energetics favours screening
is the string breaking scale $r_b$. (We shall use the term
`string breaking' when a source is screened to a different
representation, even if the latter is not the trivial one.) 
So if we calculate the potential for our sources we can expect 
the $r$-dependence to be given by 
$V_{\cal R}(r)$ for $r\leq r_b$ and $V_{{\cal R}^{\prime}}(r)$
for  $r\geq r_b$. If $r_b$ is large enough then we will
be able to extract $\sigma_{\cal R}$ from the linearly
rising potential at $r\leq r_b$.  In practice, however,
the string breaking scale is similar to other dynamical
scales in the theory and it is not clear whether any apparent
linear rise of $V(r)$ for $r \leq r_b$ is due to the
precocious formation of a string, from which we can read off
$\sigma_{\cal R}$, or if it is merely accidental. Indeed
it may be that one cannot assign an unambiguous meaning
to the quantity $\sigma_{\cal R}$ under these circumstances.
However it is also possible that if the string breaking is
relatively weak then one may be able to calculate
$\sigma_{\cal R}$ for $r\geq r_b$ by identifying an appropriate 
excited string state. In any case it is clear that string breaking
creates substantial extra ambiguities in any attempt to calculate 
the properties of strings corresponding to higher representation 
charges.

For SU(2) and SU(3) any representation $\cal R$ can be screened
by gluons to either the trivial or the fundamental representation.
However for SU($N\geq 4$) this is no longer the case and one
finds new strings that are completely stable and to which
none of the above ambiguities apply. The situation may be
summarised as follows. (We leave a fuller discussion to the
Appendix.) Suppose the representation $\cal R$ can be obtained 
from the product of $n_{\cal R}+k$ fundamental representations and
$n_{\cal R}$ conjugate ones. Let $z$ be an element of the centre, $Z_N$, 
of the SU($N$) gauge group. 
Under such a centre gauge transformation the source 
will transform as $z^k$. We shall refer to $k$ as the ${\cal N}$-ality
of the representation. Now, since gluons transform trivially
under the centre, the source will continue to transform in this 
way even if it is screened by gluons to some other representation.
Thus the same value of $k$ will label a source and all the sources
that can be obtained from it by screening. Indeed one can show
that if two representations have the same value of $k$ then either
one can be screened by gluons to the other. Within any given
class of such sources there will be a lowest string tension
$\sigma_k$, which, by string breaking, will provide the potential
for any of these sources at large enough distances. The independent
values of $k$ are constrained: under charge conjugation $k\to -k$,
and we also have $z^k = z^{N-k}$. Thus for SU($N$) we have
stable strings labelled by $k=1, ..., k_{max}$ where $k_{max}$
is the integer part of $N/2$ and $k=1$ is, of course, the fundamental
string. That is to say, we must go to at least SU(4) to have 
a $k=2$ string, and to at least SU(6) to find a $k=3$ string. 

In this paper we shall compare the $k=2$ and $k=1$ string tensions
in SU(4) and SU(5) gauge theories in D=3+1. We shall do the same
in SU(4) and SU(6) in D=2+1; and in this last case we shall
calculate the $k=3$ string tension as well. In all cases we
shall extrapolate to the continuum limit and the aim is to
obtain results that are accurate enough to distinguish between
various theoretical expectations.  Since these strings
are all stable there is no intrinsic ambiguity in defining 
a string tension and we can, in principle, achieve this goal.

We shall calculate $\sigma_k$ not from the potential between
static charges but from the mass of a $k$-string that winds
once around the spatial torus. If the string length $l$ is 
sufficiently large, its mass will be given by an expression 
similar to eqn(\ref{eqn_pot}):
\be
m_k(l) \stackrel{l\to\infty}{=} 
\sigma_{k} l - 
{{\pi(D-2)}\over{6}}{c_s\over l} + \ldots .
\label{eqn_poly}
\ee
We note that because of the different boundary conditions on
the ends of the string (periodic rather than fixed) the
$O(1/r)$ universal string correction is four times as
large as for the static potential
\cite{Polystring}.
We further note that because there are no explicit sources
there is no analogue, at small $l$, of the Coulomb potential 
in eqn(\ref{eqn_potcoul}). That is to say, this
is a particularly favourable context in which to calculate 
the string correction: its coefficient is large, and there
is no danger of confusing it with a Coulomb interaction
which has the same functional form. 

One can of course consider such closed but non-contractible 
winding
strings in any representation $\cal R$. However, just as
with the static potential, such a string can be screened to a
different string, corresponding to a representation 
${\cal R}^\prime$, as long as both strings possess the same 
$\cal N$-ality. One can picture the string breaking as follows: 
a pair of gluons pops out of the vacuum somewhere along 
the string. These then move away from each other along the
string. As they do so the section of string between
them will no longer belong to $\cal R$ but rather to
the product of $\cal R$ and the adjoint representation.
If the two gluons propagate all the way around the torus they
can meet and annihilate leaving a new string 
that is entirely in this different
representation. Clearly one can extend this to any number
of gluons. This is just like the breaking of the string 
between static sources except that here the gluons eventually 
annihilate rather than adhering to a source. Thus there is
no extra mass $\Delta M$ to consider and the breaking
can occur for small $l$ if this lowers the mass of the 
loop. That is to say, there is no region $r \leq r_b$
where one might hope to see a portion of the original string 
prior to its breaking. Of course, just as for static charges,
one might hope to see the unstable string as an excited
`resonant' string in the string mass spectrum. 

In addition to the complete string breaking described above,
the gluons may propagate only some short distance 
along the string before returning and annihilating. 
These virtual processes will renormalise
$\sigma_{\cal R}$, and simple theoretical expectations for the
string tension need to take this effect into account.

Since we are considering larger SU($N$) groups (partly in order 
to calculate $\sigma_k$ for larger $k$) one immediate question
is how this screening will depend on $N$. In particular
we know that particle decay widths vanish in the large-$N$
limit
\cite{largeN} 
and so it is natural to ask if screening will vanish in a 
similar way. The answer is yes and no. To appreciate this
consider, say, the decay $\rho \to 2\pi$
in large-$N$ QCD. This is suppressed by a factor of $1/N$.
However this suppression does not arise from the decay 
{\it per se}, but is a consequence of confinement 
constraining the pions to be colour singlets. If the
theory were not confining, so that the `$\pi$'-mesons
belonged to the adjoint representation of the colour group,
then this decay of the $\rho$ would be unsuppressed once
we summed over all the coloured $2\pi$ final states. Thus the
large-$N$ suppression of particle decays can be thought of 
as a phase-space suppression due to confinement. In just
the same way the process of gluon screening (and renormalisation)
of strings will be unsuppressed at large $N$. However the
screening of a string in representation $\cal R$ to a 
particular representation ${\cal R}^\prime$ in the same
$\cal N$-ality class may be suppressed. Whether it is or is 
not will depend on the number of states  in ${\cal R}^\prime$.
So, for example, adjoint string breaking, i.e. the adjoint sources
being screened by gluons to singlets, will be suppressed as 
$N\to\infty$. So will be the screening of $k=1$ strings down to 
the fundamental and in general the screening of $k$-strings
to the representation with $k$ quarks.
On the other hand the transformation of the mixed
to totally antisymmetric $k=2$ representations is not suppressed.
(See Appendix~\ref{a3} for details.)
Of course these general counting arguments should be supplemented
by any dynamical information we have. For example we expect
$\sigma_k \to k\sigma$ as $N\to\infty$, from the suppression
of fluctuations in that limit (and the dominance of a single
Master field). This has implications for decays as well.
\section{Lattice preliminaries}
\label{sec_prelim}

The way we perform our lattice calculations is entirely
standard and follows the pattern described in
\cite{glue01}.
For completeness we shall provide a brief summary here.

We shall work on a hypercubic lattice with periodic 
boundary conditions. The degrees of freedom are SU($N$) 
matrices, $U_l$, residing on the links, $l$, of the lattice.
In the partition function the fields are weighted with
$\exp\{S\}$ where $S$ is the standard plaquette action
\be
S = - \beta \sum_p \biggl (1 - {1\over N} ReTr \, U_p \biggr ),
\label{eqn_action}
\ee
i.e. $U_p$ is the ordered product of the matrices on
the boundary of the plaquette $p$. For smooth fields this
action reduces to the usual continuum action with
$\beta = 2N/g^2$ in D=3+1 and $\beta = 2N/ag^2$ in D=2+1 
(where $g^2$ has dimensions of mass and the theory is
super-renormalisable). By varying the inverse
lattice coupling $\beta$ we vary the lattice spacing $a$.  

The Monte Carlo we use mixes standard heat-bath and over-relaxation 
steps in the ratio $1:4$. These are implemented by updating
SU(2) subgroups using the Cabibbo-Marinari prescription
\cite{cabmar}.
We use 3 subgroups in the case of SU(3), 6 for SU(4), 10 for
SU(5) and 15 for SU(6). To check that we have enough
subgroups for efficient ergodicity
we use the same algorithm to minimise the action.
We find that with the above number of subgroups, the SU($N$) 
lattice action decreases more-or-less as effectively
as it does in the SU(2) gauge theory.
We calculate correlation functions every 5'th sweep.

We calculate correlations of gauge-invariant operators $\phi(t)$, 
which depend on field variables within a given time-slice, $t$.
The basic component of such an operator will typically be the
(traced) ordered product of the $U_l$ matrices around some closed
contour $c$.   A contractible contour, such as the plaquette itself,
is used for glueball operators. If, on the other hand, we use 
a non-contractible closed contour, which winds once around the 
spatial hyper-torus, then the operator will project onto winding
strings of fundamental flux. In the confining phase the theory is
invariant under a class of centre gauge transformations and this 
ensures that the overlap between contractible and non-contractible  
operators is exactly zero, i.e. the string cannot break. For our 
lattice action the correlation function of such an operator 
has good positivity properties, i.e. we can write
\be
C(t) = \langle \phi^{\dagger}(t) \phi(0) \rangle
= 
\sum_n | \langle \Omega | \phi | n \rangle |^2
\exp \{- E_n t \}
\label{eqn_corrln}
\ee
where $|n\rangle$ are the energy eigenstates, with $E_n$ the 
corresponding energies, and $|\Omega\rangle$ is the vacuum state. 
If the operator has $\langle \phi\rangle = 0$ then the vacuum will
not contribute to this sum and we can extract the mass of the
lightest state with the quantum numbers of $\phi$, from the large-$t$
exponential decay of $C(t)$. To make the mass calculation more 
efficient we use operators with definite momentum. 
(We will often use ${\vec p} = 0$; however, as we will see, when better
precision is required, it can be useful to extract extra information
from the smallest non-zero momenta.) Note that on a lattice
of spacing $a$ we will have
$t=a n_t$, where $n_t$ is an integer labelling the time-slices,
so that what we actually obtain from eqn(\ref{eqn_corrln}) is  
$aE_n$, the energy in lattice units.

In practice a calculation using the simplest lattice string 
operator is inefficient because the overlap onto the lightest 
string state is small and so one has to go to large values of 
$t$ before the contribution of excited states has died away;
and at large $t$ the signal has disappeared into the statistical 
noise. There are standard methods
\cite{blocking}
for curing this problem, using blocked (smeared) link operators 
and variational techniques. 
Here we use the simple blocking technique described in detail in
\cite{mtd3}.
We then have a set of trial operators corresponding to different 
blocking levels. From the space of operators spanned by these 
we can  determine the best operator using standard variational 
techniques
\cite{mtd3}.

Having determined our `best' operator,
we then attempt to fit the corresponding correlation function, 
normalised so that $C(t=0) = 1$, with a single exponential in $t$. 
(Actually a $\cosh$ to take into account the
temporal periodicity.) We choose fitting intervals $[t_1,t_2]$
where initially $t_1$ is chosen to be $t_1=0$ and then is increased
until an acceptable fit is achieved. The value of  $t_2$ is chosen
so that there are at least 3, and preferably 4, values of $t$
being fitted. (Since our fitting function has two parameters.)
Where $t_1 = 0$ and the errors on $C(t=a)$ are much smaller
than the errors at $t\geq 2a$, this procedure provides no significant
evidence for the validity of the exponential fit, and so we use the 
much larger error from  $C(t=2a)$ rather than  $C(t=a)$. (This 
typically only arises on the coarsest lattices and/or for very massive 
states.) We ignore correlations between statistical errors
at different $t$ and attempt to compensate for this both by demanding 
a lower $\chi^2$ for the best acceptable fit and by not extending
unnecessarily the fitting range. (Although in practice the error
on the best fit increases as we increase the fitting range,
presumably because the correlation in $t$ of the errors is modest 
and the decorrelation of the operator correlations is less efficient
as $t$ increases.) The relatively rough temporal 
discretisation of a few of our calculations, means that, at the
margins, there are inevitable ambiguities in this procedure.
These however decrease as $a\to 0$. Once a fitting range is chosen, 
the error on the mass is obtained by a jack-knife procedure which 
deals correctly with any error correlations as long as the binned 
data are statistically independent. Typically we take 50 bins,
each involving between 2000 and 40000 sweeps depending on the
calculation. It is plausible that bins of
this size are independent; however we have not stored our
results in a sufficiently differential form that we can
calculate the autocorrelation functions so as to test this in 
detail. A crude test is provided by recalculating 
the statistical errors using bins that are twice as large.
We find the errors are essentially unchanged when we do so, 
which provides some evidence for the statistical
independence of our original bins.

In addition to the tension of the fundamental $k=1$ string we 
also calculate tensions of $k=2$ and $k=3$ strings. Denote 
by $P_c$ the ordered product of the $U_l$ around a 
non-contractible loop $c$ that winds once around the spatial
torus. So $Tr P_c$ will project onto a winding loop of
fundamental flux. The operators  $Tr P^2_c$ and
$\{Tr P_c\}^2$ will project onto $k=2$ loops, while the
operators $Tr P^3_c$, $Tr P_c\{Tr P_c\}^2$ and $\{Tr P_c\}^3$
will project onto $k=3$ loops. These operators together with
the same ones using blocked links, are summed so as to have 
${\vec p} = 0$ and are then used as the basis of our
variational calculation for the $k=2$ and $k=3$ strings
respectively. 

We shall frequently perform fits to our extracted masses
or string tension ratios. This may be to extract
a continuum limit, or to calculate the L\"uscher string
correction. The best fit is obtained by minimising the
$\chi^2$ and the value of the latter is used to
determine whether the fit is acceptable. Equally
conventional is our estimate of the error. Suppose
we wish to calculate some quantity $A$ through the fit.
Let us suppose that the calculated `data' values are
$x_i$ with cerresponding errors $\sigma_i$. 
An estimate of the error $\sigma_A$ is provided by
\be
\sigma^2_A
=
\sum_i 
\sigma^2_i \biggl\{\frac{\partial A}{\partial x_i}\biggr\}^2 .
\label{eqn_error}
\ee
When the errors are small this formula is adequate,
but may become unreliable for larger errors and poorer
fits. It is however widely used and we therefore
adopt it to facilitate comparison with other work.

\section{A universal string correction?}
\label{sec_strings?}

Whether the long-distance dynamics of a confining flux
tube is described by an effective string theory and,
if so, what is its universality class are fundamental 
theoretical questions. These are also important practical 
questions; particularly for an accurate determination 
of the string tension, since the answer will determine 
how large is
the slowly falling $O(1/l)$ correction to the mass of a long 
flux tube in eqns(\ref{eqn_pot},\ref{eqn_poly}).
These are, however, difficult
questions to answer numerically requiring, as they do, the
accurate calculation of flux tube masses when these are
very long and very massive. Thus, although this problem has 
been addressed many times in the past, the numerical evidence 
is, as yet, far from convincing. In this Section we shall
describe some calculations which aim to improve significantly
upon this unsatisfactory situation.  

Ideally we would like to perform calculations for the
various SU($N$) gauge groups that are of interest to us in
this paper. In practice our limited computational resources
force us to focus upon the SU(2) group. Ideally, again,
we would wish to perform calculations for several values
of $a$ but again this is not practical. Instead we shall
perform calculations at a single value of $a$ which is
small enough, $a\surd\sigma \simeq 0.16$, that we can be 
confident that we are on the weak-coupling side of any 
roughening transition. We shall perform such calculations
separately for 2+1 and 3+1 dimensions since both cases
are of interest and they need not be the same. 

When is a string `long'? Since we expect
$\xi_\sigma \equiv 1/\surd\sigma$ to provide the natural length
scale for the physics of the confining flux tube, a string
of length $l=aL$ will be long if $l/\xi_\sigma = La\surd\sigma \gg 1$.
We can translate to more familiar physical units by recalling that 
in the real world $1/\surd\sigma \simeq 0.45 \mathrm{fm}$. Since quenched QCD
provides a good approximation to QCD, we can, for qualitative
purposes, use the same scale in the D=3+1 SU(3) gauge theory.
Since it appears that all D=3+1 SU($N$) gauge theories are 
`close' to each other
\cite{glue01},
it should not be too misleading to use the same scale in all
cases. For purposes of orientation (and nothing else) 
we shall also use this scale in $D=2+1$ SU($N$) gauge theories 
(where again all SU($N$) gauge theories are `close' to each other
\cite{mtd3}).
\subsection{SU(2) in 2+1 dimensions}
\label{subsec_su2d3}

We perform calculations on $L^2 L_t$ lattices at $\beta=9.0$.
(Recall that in $D=2+1$ SU($N$) gauge theories
\cite{mtd3}
the coupling $g^2$ has dimensions of mass, the theory is 
super-renormalisible, and $\beta\to 2N/ag^2$ as $a\to 0$.)
The flux tube winds around the spatial torus and so has
length $l=aL$. We perform calculations for a large number
of lattice sizes, ranging from $L=8$ to $L=40$. Recall that 
at this value of $\beta$ one has $a\surd\sigma \simeq 0.162$
\cite{mtd3}
so that the length of our flux tube ranges from 
$l=8a \simeq 1.3\times \xi_\sigma$ to
$l=40a \simeq 6.5\times \xi_\sigma$. In our `fermi' units
the latter translates to $l\simeq 3 \mathrm{fm}$. 
This should certainly be long enough 
to be governed by the long distance effective string dynamics 
if that indeed provides the correct description.

In addition to the masses, as extracted from the ${\vec p} = 0$
operators, we also calculate the energies corresponding to
the lowest five non-zero momenta transverse to the string:
$ap=2\pi n/L$ for $n=1,..,5$. If we want to use these energies 
to provide extra information on the flux loop mass,
care is needed because the continuum energy-momentum dispersion
relation, $E^2=p^2+m^2$, suffers lattice corrections. To 
determine these we have fitted the energies on our largest 
lattices with a more general dispersion relation
\be
(aE)^2 = (am)^2 + (ap)^2 + \gamma (ap)^4.
\label{eqn_Eplat}
\ee
We find that for the largest lattices the size of the lattice 
correction $\gamma$ is consistent with zero within small errors. 
For example on the $L=40$ lattice we obtain  $\gamma= 0.06(8)$
for $n \leq 5$. Thus in these cases we can simply set $\gamma=0$ 
and use the continuum dispersion relation. For smaller lattices 
the gap between momenta becomes larger and, not surprisingly,
the number of momenta that can be well fitted with $E^2=p^2+m^2$
decreases. Since these larger values of $E$ have larger 
statistical errors there is not much to be gained 
by attempting to include them and so we simply exclude
them from the fits. For the same reason we do not bother with
${\vec p} \not= 0$ for lattices with $L < 20$.

We note that the lattice correction in eqn(\ref{eqn_Eplat})
is $\gamma (ap)^4 = \gamma (2\pi n/L)^4$ and this is of the
same order as the higher order non-universal string 
corrections that we have ignored in eqn(\ref{eqn_poly}). 
We also note that the
tree-level lattice dispersion relation provides very bad
fits to our calculated energies -- it clearly contains 
lattice corrections that are far too large.
Finally we remark that if we generalise eqn(\ref{eqn_Eplat})
so as to allow a renormalisation of the $O(p^2)$ term,
we find that the fitted coefficient is unity within 
very small errors. 

We list the masses that we obtain from our $p=0$ correlators
in Table \ref{table_d3su2_stringm}. To obtain the values in
the first column we have used fits to the correlation functions, 
$C(t)$, down to the lowest plausible values of $t$, so as to 
minimise the errors (which grow with $t$). In some cases there 
are indications from the effective masses at larger $t$ that this 
mass estimate might be optimistic. So we have listed in the
second column mass estimates that we regard as very safe, but 
which might, as a result of being overcautious, overstate the 
errors and hence weaken the statistical significance of our final 
fits. Table \ref{table_d3su2_stringm} also contains, in the
third column, the mass estimates obtained by using both $p=0$ 
and $p\not= 0$ energies in the way described above. Statistically 
these are the most accurate results, although they run the risk
of possessing a small systematic bias from lattice corrections
to the continuum energy-momentum dispersion relation. However
any such bias will be smallest on the largest lattices and
it is only on these that the $p\not=0$ energies make a
significant difference.

An immediately striking feature of the listed masses is that they
rise more-or-less linearly with length, all the way out to the 
longest loops. This demonstrates directly that in D=2+1 SU(2) gauge 
theories we have linear confinement out to at least $l \sim 3\mathrm{fm}$.
However this is no more than expected and so we shall not dwell
upon it any further here.

We turn now to the real question of interest here: how accurately 
can we test the $O(1/l)$ string correction term in 
eqn(\ref{eqn_poly})? As a first step we calculate the effective 
value of the coefficient $c_s$ that one obtains from pairs of
flux loops of length $l$ and $l^{\prime}$ respectively:
\be
c_s^{eff}(l,l^{\prime})
=
{{6}\over{\pi (D-2)}} \times
{
\bigl\{ {{m(l)}\over{l}} - {{m(l^{\prime})}\over{l^{\prime}}}
\bigr\} 
\over
\bigl\{ {1\over{l^{\prime 2}}} - {1\over{l^{2}}}
\bigr\} 
}
\label{eqn_polypair}
\ee
In Table~\ref{table_d3su2_cstring1} we list the values of
$c_s^{eff}(l,l^{\prime})$ that one obtains for neighbouring
values of $l$ and $l^{\prime}$ using the masses listed in
Table \ref{table_d3su2_stringm}. Any given value gives us no
information on the validity of the $O(1/l)$ string correction.
However if we find that $c_s^{eff}(l,l^{\prime})$ has a 
finite non-zero limit as $l,l^{\prime} \to \infty$ then we
will have shown that the leading correction is indeed of this
form and the limiting value will provide us with an estimate 
of the coefficient $c_s$. 

Our most accurate values of $c_s^{eff}$ are those that 
incorporate $p\not= 0$ energies, and we display these in 
Fig.\ref{fig_d3cspair}. We see from the plot that for small loop 
lengths the value of  $c_s^{eff}$ is small and increases as the 
loop length grows. However the behaviour is not monotonic: 
at intermediate $l$ the value of $c_s^{eff}$ increases through
the bosonic string value and perhaps peaks close to the 
value for a Neveu-Schwartz string. This occurs at a loop
size $l \sim 1 \mathrm{fm}$ which is typically the longest loop for which 
older calculations had usefully accurate results. Thus, by 
focusing on slightly different intervals close to $l \sim 1 \mathrm{fm}$ 
it is possible to either confirm or contradict the bosonic
string value; but, in either case, incorrectly. To obtain real 
evidence one must go to longer
strings and if one does so, as in Fig.\ref{fig_d3cspair}, one
finds that the value of $c_s^{eff}$ decreases again to
something that appears consistent with the bosonic string value.

To obtain our estimate of $c_s$ we use eqn(\ref{eqn_poly}) to 
fit all the loop masses that are longer than some reference 
value $l_0$. If $l_0$ is small there is no acceptable fit. 
As we increase $l_0$ eventually the fit becomes acceptable. 
We can then increase $l_0$ to check the stability of the best 
fit. In Table~\ref{table_d3su2_cstring2} we list the results 
of this procedure for each of our three sets of loop masses.
The most accurate values are obtained from the last set,
and are plotted in Fig.\ref{fig_d3csall}. 
From the second column of Table~\ref{table_d3su2_cstring2} 
we extract a `safe' estimate for the string correction:
\be
c_s = 1.066 \pm 0.036.
\label{eqn_csd3}
\ee
This is close to the bosonic string value and far from that
of other simple string theories.

The above analysis assumes that the leading correction
to the linear dependence of the mass is $\propto 1/l$.
The fact that $c_s^{eff}$ becomes independent, within errors, 
of $l$ once $l\geq 16a$, tells us that our results
are certainly consistent with such a string-like 
correction. It is interesting to ask how well our calculations 
exclude other choices. We have therefore performed fits
to
\be
m(l) =
\sigma l + {c\over{l^p}}.
\label{eqn_power}
\ee
We find that the there are no acceptable fits if any values
with $l<14a$ are included. The fit to $l\geq 14a$ has
a mediocre but acceptable $\chi^2$, and we find that the
possible range of powers is $p=1.4 \pm 0.5$. For $l\geq 16a$ 
the best fit is very good and one finds  $p=0.9 \pm 0.5$.
Fits to  $l\geq 20a$ are equally good but no longer provide
a useful constraint on $p$. In short, our results are 
consistent with the $O(1/l)$ string-like correction
term and, in any case, the power of $1/l$ is constrained
to be within the range $p=0.9 \pm 0.5$. So if we constrain
the power $p$ to be an integer, we find that it indeed
has to be $p=1$.

\subsection{SU(2) in 3+1 dimensions}
\label{subsec_su2d4}

We perform calculations on $L^3 L_t$ lattices at $\beta=2.55$.
At this value of $\beta$ one has $a\surd\sigma \simeq 0.159$
\cite{glue01},
so that the size of $a$ is very similar to that in our $D=2+1$
calculations. The calculations in D=3+1 are, of course,
slower and so our range of lattice sizes and our statistics
is somewhat less. One may hope that this will be partly
compensated for by the fact that the expected string correction,
$(D-2)\pi c_s/6L$, will be twice as large (for a given
universality class). We perform calculations on lattices
ranging from $L=8$ to $L=32$. Thus our longest flux loop
is $l=32a \simeq 2.3 \mathrm{fm}$ which, if our experience in $D=2+1$ 
is relevant, should be long enough to see the 
leading correction.

Our calculation and analysis is precisely as in D=2+1,
except that the values of the momenta transverse to the flux 
loop that we use are $p^2 = 0,1,2,4$. The mass estimates are
listed in Table~\ref{table_d4su2_stringm}. As in the D=2+1
case we list two sets of masses extracted from the $p=0$ 
correlators. In general our mass estimates are chosen to be
those with the smallest errors while still giving plausible fits. 
In some cases the plausibility is less than convincing and 
we then also select a `safer' mass estimate, which will
have larger errors. The former numbers provide the first
column of masses in Table~\ref{table_d4su2_stringm}
while the latter provide the second column. The two columns
only differ in some cases. The mass estimates obtained
using $p\not=0$ as well as $p=0$ correlators are also
divided into two sets. (A division that did not appear useful
in D=2+1.) The first set (third column of masses) differs
from the second principally in that on the $L=14,16$ lattices
we chose less plausible $p=0$ masses in order to be 
consistent with the $p=1$ values with which they were then
averaged. In the last `safe' column we dealt with this
discrepancy by not using the $p\not=0$ values (which, in
any case, become much less useful on the smaller lattices).
Thus this range of mass estimates gives some indication 
of any systematic error that arises from our procedure
for extracting masses. 

We first note that the loop mass increases approximately
linearly with the loop length confirming, as expected,
that the theory is linearly confining (up to $\sim 2.3\mathrm{fm}$). 
 
In Table~\ref{table_d4su2_cstring1} we list the values 
of the effective string correction coefficient,
$c_s^{eff}$, defined in eqn(\ref{eqn_polypair}). As in D=2+1
our most accurate values of $c_s^{eff}$ are those that 
incorporate $p\not= 0$ energies, and we display one set of these 
in Fig.\ref{fig_d4cspair}. We see in the plot a behaviour
similar to what we observed in D=2+1: the value of  $c_s^{eff}$
increases as the loop length increases, attains a maximum
value at $l\simeq 1 \mathrm{fm}$ that is significantly larger than 
the bosonic string value, and then decreases to a value consistent
with the value for a bosonic string. 

Just as in D=2+1 we  estimate $c_s$ by using  eqn(\ref{eqn_poly})
to fit all the loop masses that are longer than some reference 
value $l_0$. If $l_0$ is small there is no acceptable fit. 
As we increase $l_0$ we find that the fit 
eventually becomes acceptable. 
We then increase $l_0$ to check the stability of the best 
fit. In Table~\ref{table_d4su2_cstring2} we list the results 
of this procedure for each of our last three sets of loop masses.
(For the first set there are no acceptable fits, perhaps  
indicating that some of the mass choices were indeed
too optimistic.)
The most accurate values are obtained from the last two sets,
and we use these to obtain our
best estimate for the string correction:
\be
c_s = 0.94 \pm 0.04.
\label{eqn_csd4}
\ee
This is close to the bosonic string value and far from that
of other simple string theories.

\section{$k$-strings}
\label{sec_kstrings}

In the previous Section we accumulated some evidence that 
SU(2) flux tubes in the fundamental representation are  
described by an effective bosonic string theory at large
distances. In this Section we consider flux tubes in
higher representations: the $k$-strings described in
the Introduction. We will not be able to perform 
comparable checks on the stringy nature of these flux
tubes although we will perform some crude finite volume
analyses that are primarily designed to confirm the presence 
of linear confinement in SU($N\geq 4$) gauge theories.
In our analysis of the string tension ratios, 
$\sigma_k/\sigma$, we shall make the plausible assumption 
that the leading correction is that of a simple bosonic string. 
However for completeness we shall also point out how the 
results are affected if this should not be the case.

In order to provide some theoretical context within which
to view our numerical results, we shall first briefly
summarise some of the existing ideas about how such ratios 
might behave. This is not intended to be an exhaustive review,
and our references are merely designed to provide an entry 
into the literature rather than aiming at completeness. 

We then describe our calculations of $k=2$ and  (fundamental)
$k=1$ strings in D=3+1 SU(4) and SU(5) gauge theories.
We follow this with a description of our D=2+1 calculations
which are for SU(4) and SU(6). In this last case we also have
non-trivial $k=3$ strings that we are able to analyse.

We shall find that in D=3+1 the string tension ratios 
lie between the predictions of MQCD and Casimir scaling,
straddling both within two standard deviations; with SU(4) slightly
favouring Casimir scaling and SU(5) leaning towards MQCD. 
In D=2+1 our results are again close to both Casimir 
scaling and to MQCD, but now they are much more accurate and 
so we can begin to see significant deviations.
Although we see deviations from both sets of predictions,
those from MQCD are much larger than those from  Casimir scaling.
We point out that near-Casimir scaling occurs 
naturally if the confining flux tube has a cross-section
that is nearly independent of the flux carried. We perform explicit
calculations in D=2+1 that suggest that this 
is in fact so. These calculations give us, as a side-product,
the value of the $k$-string tensions at high temperature,
and we find near-Casimir scaling there as well. Motivated
by this we perform a high $T$ calculation in the D=3+1 SU(4)
gauge theory where we again find near-Casimir scaling. 

It is interesting to ask if all this also occurs for the
unstable strings. We shall show that our D=2+1 calculations 
provide some evidence that points to this .

\subsection{Some expectations for ${\bf k}$-strings}
\label{subsec_expect}

The interest in strings that emanate from sources in
higher representations goes back a long way. The
early discussions were framed in terms of 
unstable strings in SU(2) and SU(3) gauge theories
as were the lattice calculations. (See, for example,
\cite{oldcs}.)
Despite the uncertainties of this kind of calculation, these 
early results were already seen as being able to discriminate
against particular theoretical ideas; in particular 
\cite{bagadj}
against the bag model
\cite{bagadj,bag}.
There have been recent much more accurate SU(3) calculations
\cite{deldar1,bali}
that support this earlier work, and this has sparked some
interest in the possible dynamics
\cite{deldar2,simonov,suzuki}.
The recent interest
\cite{winoh,string00},
including our own, in stable $k$-strings in SU($N\geq 4$) 
has been largely due to conjectures arising in M(-theory)QCD 
\cite{mqcd}. 
Here we briefly allude to some of these theoretical ideas, with
a particular focus on MQCD and `Casimir scaling' since their 
predictions turn out to be closest to the results we
obtain 
\cite{string00,glue01}
for the tensions of $k$-strings.

\subsubsection{unbound strings}
\label{subsubsec_unbound}

The simplest possibility is that 
\be
\sigma_k = \tilde{k} \sigma_{k=1} 
\ \ \ \ \ \ ;  \ \ 
\tilde{k} = \mathrm{min}\{k,N-k\}.
\label{eqn_unbound}
\ee
That is to say the total flux is carried by $k$ (or $N-k$
if that is smaller) independent fundamental flux tubes. 
This would occur if the interaction between fundamental flux
tubes was so weak that there were no bound or resonant 
multi-string states. One may regard this as the trivial 
scenario with which the actual calculated values of 
$\sigma_k$ can be contrasted.

\subsubsection{Casimir scaling}
\label{subsubsec_casimir}

The idea that the confining flux tube between sources
will be proportional to the quadratic Casimir, ${C_{\cal R}}$,
of the representation of those sources
\be
\sigma_{\cal R} \propto  C_{\cal R}
\label{eqn_casimirR}
\ee
is an old idea. An early motivation
\cite{oldcs}
arose from a model of `random fluxes' for the vacuum
and the observation that in certain solid state systems
this leads to a dimensional reduction $D \to D-2$. Thus
$D=4$ theories would reduce to $D=2$ gauge theories in
which the Coulomb linear potential is indeed 
proportional to the quadratic Casimir. The numerical
calculations supporting this were in D=4 SU(2)
\cite{oldcs}
and involved potentials at relatively short distances.
The observation soon after that one seemed to see a
similar Casimir scaling in D=3 theories
\cite{oldcs}
forced a generalisation of the dimensional reduction idea
\cite{oldcs}.
The most accurate early calculations involved the adjoint
string tension. Recently however there have been accurate
calculations 
\cite{deldar1,bali}
for a variety of representations in SU(3) and this has
sparked renewed interest in this idea
\cite{deldar2,simonov,suzuki}.

There are obvious ambiguities in calculating the string
tension of unstable strings from the intermediate distance
behaviour of the static potential. At short distances we
know that we have a Coulomb potential which, of course,
displays Casimir scaling. As the potential interpolates
between this and the long-distance behaviour one expects
some continuity. If, as is usually done, one fits the potential 
$V(r)$ by a simple sum of a Coulomb term and a linear piece,
$V(r)= V_0 + c/r + \sigma r$, and then performs the fit in a
limited range of $r$ immediately beyond the Coulomb region, 
then it might be that simple continuity artificially forces 
approximate Casimir 
scaling on the fitted linear term. While this is no more than
a possibility, it does underscore the utility of using
stable strings, as we shall do, where one can go to larger 
distances, and
doing the calculation in a way, as we shall also do, that does
not involve explicit sources and associated Coulomb terms.

Since the sources may be screened by gluons, which are 
in the adjoint representation and do not feel centre gauge
transformations, it is appropriate, as we remarked
earlier, to categorise
the representations of SU($N$) sources by how they transform 
under the centre of the group. If the source acquires a factor
$z^k$ with $k\in Z_N$, then we shall generically refer to
the corresponding flux tube as a $k$-string. Any $k$-string
can be transformed into any other $k$-string by appropriate
gluon screening. Thus the stable $k$-string will be the 
one with the smallest string tension. Any other $k$-string
will, at sufficiently large distances, find it energetically
favourable to be transformed into the lightest stable string
through gluon screening. If we have Casimir scaling then
the lowest string tension corresponds to the representation
with the smallest quadratic Casimir, and this is the totally
antisymmetric representation. (See the Appendix~\ref{a2}.) 
The ratio of these quadratic Casimirs gives
us the Casimir scaling prediction for  stable $k$-strings 
\be
{{\sigma_{\mathit{k}}}\over{\sigma}}
= 
{{k(N-k)}\over{N-1}}.
\label{eqn_casimir}
\ee

\subsubsection{MQCD}
\label{subsubsec_mqcd}

A number of calculations in  brane (M-)theory of QCD-like
theories (see
\cite{mqcd} 
and references therein), which are generically referred to as 
MQCD, find that that the string tension of $k$-strings satisfies 
\be
{{\sigma_{\mathit{k}}}\over{\sigma}}
=
{{\sin \frac{k\pi}{N}}\over{\sin \frac{\pi}{N}}} .
\label{eqn_mqcd}
\ee
This led to the conjecture 
\cite{mqcd} 
that this might be a universal result and that QCD (and SU($N$) 
gauge theories) fall into this universality class.

This prediction has reasonable properties: it has the
required $k \leftrightarrow N-k$ symmetry and takes
sensible values for $N=2,3$. However the MQCD derivation
neglects potentially important quantum fluctuations which might
\cite{mqcd} 
renormalise the simple and elegant formula in eqn(\ref{eqn_mqcd}).

The MQCD calculations are, strictly speaking, for SU($N$) gauge 
theories in 3+1 dimensions. It is not clear how much evidence
there would be for a corresponding MQCD conjecture in D=2+1, 
although a naive reading suggests that the brane constructions in
\cite{mqcd} 
would lead to the same conclusion for the $\sigma_k$ ratios.
In any case the trigonometric formula in eqn(\ref{eqn_mqcd})
has the correct qualitative properties and so we shall
compare our results to it not only in D=3+1 but also in
D=2+1 and, indeed, at finite temperature.

\subsubsection{bag model}
\label{subsubsec_bag}

In the bag model (see e.g.
\cite{bag,bagadj})
the flux between distant sources is confined to a cylindrical
bag of cross-section $A$. The flux is homogeneous,
\be
E_a A = g T_a ,
\label{eqn_flux}
\ee
and the vacuum energy difference between the 
inside and outside of the bag is given by the bag constant $B$.
Thus the energy per unit length is 
\cite{bag,bagadj}
\be
{E \over l} 
=
2\pi\alpha_s {{C_{\cal R}}\over A} +A B
\label{eqn_bagV}
\ee
where ${C_{\cal R}}$ is the quadratic Casimir of the source
and $\alpha_s$ is the strong coupling constant.
One now fixes the area $A$ by minimising the energy. This
gives the string tension to be
\be
\sigma_{\cal R} \propto  \{C_{\cal R}\}^{\frac{1}{2}}
\label{eqn_bag}
\ee
which differs markedly from Casimir scaling. The fact
that the early numerical calculations gave an adjoint string
tension that satisfied eqn(\ref{eqn_casimirR}) rather
than eqn(\ref{eqn_bag}) was picked up
\cite{bagadj}
as providing critical evidence against conventional bag dynamics
\cite{bag,bagadj},
in that it suggested a flux tube
cross-section that was independent of the size of the flux.

\subsubsection{strong coupling}
\label{subsubsec_gstrong}

In the strong coupling limit, $\beta \to 0$, of our action,
a Wilson loop involving $k$ strings will need to be tiled with 
plaquettes at least $k$ (or $N-k$) times. The leading term
in this limit will therefore reproduce eqn(\ref{eqn_unbound}):
$\sigma_k = \tilde{k} \sigma_{k=1}, \tilde{k} =
\mathrm{min}\{k,N-k\}$. However the non-leading terms will 
introduce interactions between these tiled surfaces, and this
simple ratio will change as we move away from the strong 
coupling limit.

Strong coupling predictions are, of course, not universal;
however this one is more universal than most. If we
generalise the action to contain any combination of closed
loops, so long as these are linear in the SU($N$) link 
matrices we will still obtain eqn(\ref{eqn_unbound}). 
However if we include loops or products of loops that
are not linear in the links then we can obtain other
results. One can think of the action as having loops
in different representations, and the value of $\sigma_k/\sigma$
will depend only on what these representations are and what are
their relative weights. By choosing an action in an
appropriate `universality' class, one can essentially obtain
for $\sigma_k/\sigma$ any value one wants. 

Hamiltonian strong coupling (see e.g.
\cite{books})
is more interesting. The leading
term, as $g^2\to\infty$, is simply the quadratic Casimir 
for each spatial link. Gauss's law means that our two 
$k$-sources are joined by excited links, and that the
lightest $k$ string will satisfy Casimir scaling as in
eqn(\ref{eqn_casimir}). Of course the magnetic perturbation
will spoil this result as we move away from $\beta=0$.

\subsection{$k$-strings in D=3+1}
\label{subsec_d4}

We will now calculate the ratio of the $k=2$ and fundamental
string tensions, $\sigma_{k=2}/\sigma$, in both SU(4) and
SU(5) gauge theories. There are no other stable $k$-strings
for these values of $N$ but having results for two values
of $N$ will already provide significant constraints.
We are, of course, interested in the continuum
limit, so we calculate this ratio for several lattice 
spacings and then extrapolate to the continuum limit using
the fact that for the plaquette action the leading lattice
correction to dimensionless mass ratios is $O(a^2)$:
\be
{{\sigma_k(a)} \over {\sigma(a)}} =
{{\sigma_k(0)} \over {\sigma(0)}} + c a^2\sigma .
\label{eqn_extrap}
\ee
We calculate the string tension from the mass of a flux
loop that winds around the spatial torus. We assume
that the leading correction to the linear dependence
of the mass is that appropriate to a simple bosonic string: 
\be
m_k(l) \stackrel{l\to\infty}{=} 
\sigma_{k} l - 
{{\pi(D-2)}\over{6}}{1\over l} .
\label{eqn_bosestring}
\ee
This assumption has some support from the calculations
of the previous Section, but it is not guaranteed that
what holds for SU(2) holds also for larger $N$. So
we shall occasionally pause to state how sensitive are
our results to this assumption.

We begin by listing in Table~\ref{table_scaled4}
the (fundamental) string tensions 
\cite{glue01}
corresponding to the various $\beta$ values at which we
perform our calculations. This sets the scale of $a$
in physical units. In Table~\ref{table_datsu4d4}
we list our lattices and calculated values of the
$k=1$ and $k=2$ flux loop masses for the case of SU(4);
and in Table~\ref{table_datsu5d4} for SU(5). (Note that
in these calculations we only  use $p=0$ correlators.)

In order to extract a string tension from the flux loop
mass we must ensure that our loop length is long enough
for the corrections to eqn(\ref{eqn_bosestring}) to be
negligible within our statistical errors. In
Section \ref{subsec_su2d4} we have seen that in the case
of SU(2) this appears to be the case for strings longer 
than $l\surd\sigma \equiv La\surd\sigma \simeq 3$ (see
Tables~\ref{table_d4su2_cstring1},\ref{table_d4su2_cstring2}).
Here we perform an additional finite size study, this time
in SU(4), which, while less accurate, will probe the behaviour 
of $k=2$ as well as $k=1$ strings. 

Our finite size study is at $\beta=10.7$ and involves loops
ranging from $L=6$ to $L=16$ with masses as listed in
Table~\ref{table_datsu4d4}. The longest length translates
into $l \simeq 4.9 /\surd\sigma \simeq 2.2 \mathrm{fm}$. We observe
that both the $k=2$ and $k=1$ masses grow approximately
linearly with $l$,  demonstrating that the SU(4) theory
linearly confines both $k=1$ and $k=2$ charges (at least 
over this distance range). Using eqn(\ref{eqn_bosestring})
we extract the ratio  $\sigma_{k=2}/\sigma$ which we
plot in Fig.\ref{fig_d4su4string}. We see that within 
errors the ratio becomes independent of the flux loop
length for $l\geq 10a \simeq 3/\surd\sigma$. For comparison
we also show what happens if we do not include any 
string correction at all, i.e.  $\sigma_{k=2}/\sigma 
= m_{k=2}(l)/m_{k=1}(l)$. We see that while the ratio changes 
by a few percent, it becomes independent of $l$, within errors,
at the same length, $l=10a$. By the same token it is clear
that the results of this calculation are not accurate enough
to distinguish between different possible string corrections.

Our finite volume study has taught us that higher order
corrections in $1/l$ to the string tension ratio will
be negligible (within our typical errors) if we make
sure that our loop length satisfies $l\surd\sigma\geq 3$.
Comparing the values of $a\surd\sigma$ in  
Table~\ref{table_scaled4} with the corresponding lattice
sizes listed in Table~\ref{table_datsu4d4} and
Table~\ref{table_datsu5d4} we see that our loop lengths
have been chosen to fulfill this bound; more generously
at smaller $a$ where the errors are smaller.

Assuming eqn(\ref{eqn_bosestring}), we extract our string
tension ratios and plot them against $a^2\sigma$ in 
Fig.~\ref{fig_d4sig2}. (At $\beta=10.7$ we use only the
$L=10,12$ lattices since the larger volumes have errors
that are too large to be useful.) On such a plot the continuum 
extrapolation, eqn(\ref{eqn_extrap}), is a simple
straight line. We show the best such fits in Fig.~\ref{fig_d4sig2}.
We find that if we use all the points we get an excellent $\chi^2$ 
for SU(4) and an acceptable one for SU(5). From these fits
we obtain the following continuum values:
\be
\lim_{a \to 0}
{{\sigma_{\mathit{k=2}}} \over {\sigma}}
=
\left\{ \begin{array}{ll}
1.357 \pm 0.029  & \ \ \ {\mathrm{SU(4)}} \\
1.583 \pm 0.074  & \ \ \ {\mathrm{SU(5)}}
\end{array}
\right. .
\label{eqn_contd4}
\ee
One might worry that these fits could be biased by including 
the coarsest $a$ value (where we know 
\cite{glue01}
the lattice corrections to the scalar glueball mass to be
large). If we exclude this coarsest $a$ point our best
values in eqn(\ref{eqn_contd4}) are changed to:
\be
\lim_{a \to 0}
{{\sigma_{\mathit{k=2}}} \over {\sigma}}
=
\left\{ \begin{array}{lll}
1.377 \pm 0.035  & \ \ \ \beta\geq 10.70  & \ \  {\mathrm{SU(4)}} \\
1.76  \pm 0.14   & \ \ \ \beta\geq 16.975 & \  \ {\mathrm{SU(5)}}
\end{array}
\right. .
\label{eqn_contd4b}
\ee
This gives us some idea of the direction of any such bias.

It is interesting to compare our results with the 
expectations of MQCD
\be
{{\sigma_{\mathit{k=2}}}\over{\sigma}}
\stackrel{MQCD}{=}
{{\sin \frac{2\pi}{N}}\over{\sin \frac{\pi}{N}}} 
=
\left\{ \begin{array}{ll}
1.41 ... & \ \ \ {\mathrm{SU(4)}} \\
1.61 ... & \ \ \ {\mathrm{SU(5)}}
\end{array}
\right. 
\label{eqn_sig2mqcd}
\ee
and Casimir scaling
\be
{{\sigma_{\mathit{k=2}}}\over{\sigma}}
\stackrel{CS}{=}
{{k(N-k)}\over{N-1}}
=
\left\{ \begin{array}{ll}
1.3{\bar 3} & \ \ \ {\mathrm{SU(4)}} \\
1.50 & \ \ \ {\mathrm{SU(5)}}
\end{array}
\right. .
\label{eqn_sig2cs}
\ee
We see that our results in eqn(\ref{eqn_contd4}) and 
eqn(\ref{eqn_contd4b}) are consistent,
at the $2\sigma$ level, with both these 
expectations, within quite small errors; with perhaps a 
slight bias towards favouring MQCD. It is because the 
two sets of predictions are numerically very similar that 
we cannot, at present, choose between them. On the other 
hand we clearly exclude the  unbound string value of 2:  
i.e if we do wish to think of the $k=2$ string 
as being composed of two $k=1$ strings then
it must be a tightly bound state of such strings. We
also clearly exclude the bag model prediction: 
\be
{{\sigma_{\mathit{k=2}}}\over{\sigma}}
\stackrel{Bag}{=}
\sqrt{{k(N-k)}\over{N-1}}
=
\left\{ \begin{array}{ll}
1.15... & \ \ \ {\mathrm{SU(4)}} \\
1.22... & \ \ \ {\mathrm{SU(5)}}
\end{array}
\right. .
\label{eqn_sig2bag}
\ee
In order to distinguish clearly between MQCD and Casimir scaling 
we need to reduce our statistical errors by about a factor
of two; a feasible goal but one for the future. 
 
Thus our conclusions are essentially unchanged from those of 
our earlier paper
\cite{string00}
although our SU(4) calculation now has smaller statistical
errors, and our SU(5) calculation is now free of the
potentially large systematic errors that concerned us 
earlier.

A final remark.
Our above analysis assumed that the flux tubes behave like 
simple bosonic strings. What difference does it make
if we do not make this assumption? Suppose first that 
we use the result we obtained in 
Section~\ref{subsec_su2d4} for the coefficient of the
string correction: $c_s=(1.25\pm0.25)\pi/3$ (where we take
very generous errors). Repeating our analysis with such
a string correction we find that our results in 
eqn(\ref{eqn_contd4}) are lowered by about 10\% of the
statistical error; that is to say, insignificantly.
Even if we were to ignore what we knew and simply
assumed some range like $c_s=(1 \pm 1)\pi/3$
we would find that the maximum shift would be less than
our quoted statistical error. (For  $c_s=0$ the ratios
are close to MQCD while for $c_s=2\pi/3$ they drop very
close to Casimir scaling.)

\subsection{$k$-strings in D=2+1}
\label{subsec_d3}

Our calculations in D=2+1 follow the same pattern
as in D=3+1 except that our calculations are in SU(4) 
and SU(6). The main reason for SU(6) rather than SU(5) 
is that with the former one also has stable k=3 strings
that one can study. On the other hand the calculations
take longer which is why we contented ourselves with
SU(5) in four dimensions. Our calculations are
summarised in Tables~\ref{table_datsu4d3} and
\ref{table_datsu6d3} and the scale of $a$, in units of
the fundamental string tension, is given in 
Table~\ref{table_scaled3}.

We begin with a finite volume study in SU(4) at
$\beta=28.0$ that parallels our D=3+1 study. The loop
lengths range from $L=4$ to $L=16$, with the largest loop 
corresponding to $l=16a\simeq 4/\surd\sigma \simeq 1.6 \mathrm{fm}$.
We extract $\sigma_{k=2}/\sigma$ using  eqn(\ref{eqn_bosestring})
and plot the result in Fig.\ref{fig_d3su4string}. We see that
the ratio of string tensions is independent of the loop
length (within errors) once
$l\geq 10a\simeq 2.5/\surd\sigma \simeq 1.1 \mathrm{fm}$.
This is a somewhat shorter length than the one we found
in D=3+1. We shall later see that the flux tube is thinner
(in units of $\sigma$) in D=2+1 than in D=3+1 and this is
presumably why the corrections are smaller.
We also show in Fig.\ref{fig_d3su4string} the string tension 
ratios one obtains
if one assumes no correction. This also plateaus for
$l\geq 10a$. Moreover we see that the value of the ratio
differs by only about 1\%. (Note the string correction
is $\propto (D-2)$ and so is larger in D=3+1 than in D=2+1.)

We observe that the loop lengths we shall use, as listed
in Tables~\ref{table_datsu4d3} and \ref{table_datsu6d3},
satisfy the above bound, $l\geq  2.5/\surd\sigma$, by a
good margin. So assuming eqn(\ref{eqn_bosestring}), 
we plot our string tension ratios in Fig.~\ref{fig_d3sigk}
against $a^2\sigma$. We also show the straight line
continuum extrapolations, using eqn(\ref{eqn_extrap}).
We find that we get an acceptable $\chi^2$ using all the
points. We thus obtain the following continuum values:
\be
\lim_{a \to 0}
{{\sigma_{\mathit{k=2}}} \over {\sigma}}
=
\left\{ \begin{array}{ll}
1.3548 \pm 0.0064  & \ \ \ {\mathrm{SU(4)}} \\
1.6160 \pm 0.0086  & \ \ \ {\mathrm{SU(6)}}
\end{array}
\right. 
\label{eqn_contk2d3}
\ee
and 
\be
\lim_{a \to 0}
{{\sigma_{\mathit{k=3}}} \over {\sigma}}
=
1.808  \pm 0.025   \ \ \ {\mathrm{SU(6)}}.
\label{eqn_contk3d3} .
\ee
We note that the errors here are much smaller than in
eqn(\ref{eqn_contd4}) and the MQCD expectation is
excluded. We also see, in our $k=2$ SU(4) ratio, a
deviation from Casimir scaling at the level of about 3.5
standard deviations. In SU(6) the expectations are
\be
{{\sigma_{\mathit{k=2}}}\over{\sigma}}
\stackrel{SU(6)}{=}
\left\{ \begin{array}{ll}
1.73... & \ \ \ {\mathrm{MQCD}} \\
1.60    & \ \ \ {\mathrm{CS}}
\end{array}
\right. 
\label{eqn_sig2su6}
\ee
and 
\be
{{\sigma_{\mathit{k=3}}}\over{\sigma}}
\stackrel{SU(6)}{=}
\left\{ \begin{array}{ll}
2.0     & \ \ \ {\mathrm{MQCD}} \\
1.8     & \ \ \ {\mathrm{CS}}
\end{array}
\right. .
\label{eqn_sig3su6}
\ee
We see that in this case our results are far from the MQCD values 
and agree well with Casimir scaling.

Given our very small statistical errors one might worry
that our assumption of a bosonic string correction might 
introduce a relatively large systematic error. In fact 
this is not so. If we take our SU(2) string analysis in 
Section~\ref{subsec_su2d3} to be telling us that the
coefficient of the $1/l$ term is $c_s=(1.1\pm 0.1)\pi/6$,
then we find a negligible shift in our above predictions.
Even if we were to assume $c_s=0$ this would shift
our values upwards by less than 2\%. While this would
increase the deviation from Casimir scaling, it would be
far from enough to bridge the gap to the MQCD prediction.
However  $c_s=0$ is a contrived and extreme example, and 
it appears to us that any reasonable estimate of the
systematic error arising from the uncertainty in the
string correction shows it to be negligible.

\subsection{Width of $k$-strings}
\label{subsec_width}

We have seen that in both D=2+1 and D=3+1 the ratio $\sigma_k/\sigma$ 
is close to the Casimir scaling value. In D=3+1 it is also
consistent with MQCD, but the MQCD and Casimir scaling predictions
are in fact quite close. If one imagines modelling the flux tube
then this is a somewhat conter-intuitive result. It would be natural
to think of the flux as homogeneous,
as in eqn(\ref{eqn_flux}), and that the vacua inside and
outside the flux tube differ by some energy density $\delta E_v$.
These are of course the ideas embodied in the Bag model. One would
then expect that as the representation of the flux increases,
so that the chromoelectric energy density increases, the area 
will increase so as to minimise the total energy increase.
This is just the variational calculation of the naive bag model
which leads to a ratio $\sigma_k/\sigma$ that grows as
the square root of the quadratic Casimir. This is definitely
excluded by our calculated values. One might imagine extending
this simple-minded model by providing the flux tube with
a surface tension. However this would have no effect in
D=2+1 where the `surface' is independent of the flux tube width.

If the flux is homogeneous and if the flux tube width is independent
of the total flux carried, then one naturally obtains Casimir scaling.
If one considers a superconductor, in a phase that exhibits
the Meissner effect and supports (magnetic) flux carrying flux tubes,
there is a range of parameter values, called the deep-London limit,
where the flux tube cross-section is indeed independent of the
flux (see e.g.
\cite{SCbook}).
This corresponds to a penetration depth, related to the photon mass,
being much larger than the inverse scalar Higgs mass. The deviation
from Casimir scaling would be related to the ratio of these
masses.

It is interesting to test these ideas. Here we shall attempt
to calculate the widths of the flux tubes corresponding to
different $k$-strings and see how close the width is to
being independent of $k$. We shall do so in D=2+1 because
it is faster; but the same technique can
be used in D=3+1. We shall perform calculations for $k=1,2$
flux tubes in SU(4) and for $k=1,2,3$ flux tubes in SU(6).

We use a technique that was employed in 
\cite{MTwidth}
to calculate the width of SU(2) flux tubes in D=2+1.
Consider a lattice of size $L\times L_{\perp} \times L_t$.
$L_t$ refers to the Euclidean time in which we calculate 
correlations. The flux loop is of length $L$, and $L_{\perp}$
is the spatial size transverse to the flux tube. By
reducing  $L_{\perp}$ we can squeeze the flux tube. If the
flux tube oscillates with simple harmonic modes then it
will not be affected by reducing the finite (periodic) 
transverse width until it reaches the `intrinsic' width
of the flux tube
\cite{MTwidth}.
Once $L_{\perp}$ is smaller than this width, which we
shall call $l_w = aL_w$, we expect the mass 
of the flux loop to increase as
\cite{MTwidth}
\be
am(L;L_{\perp})
=
am(L; \infty) \times { {L_w}\over{L_{\perp}} } \ .
\label{eqn_mwidth}
\ee
The onset of the increase is at $L_{\perp} = L_w$
and provides us with an estimate of $L_w$. Our main
interest is to see if $L_w$ varies with $k$ or not.

Of course the above arguments are very simple. There
are also changes in the vacuum as  $L_{\perp}$ becomes
small which we have neglected. In particular
there is a phase transition at a critical 
value of $L_{\perp}$ (see Section~\ref{subsec_highT}) which
is characterised by the development of a non-zero vacuum 
expectation value for the Polyakov loops that wind around 
the short $L_{\perp}$ torus. However the string tensions
we calculate behave smoothly through this transition,
suggesting that our simple analysis should not be invalidated.
In any case, our main conclusion, that as $k$ grows
the flux tube width does not grow ever larger, will
remain valid since such a growth would mean that higher $k$
flux tubes would begin to be squeezed when $L_{\perp}$
was greater than its critical value, and this would certainly 
be visible.

We show in Table~\ref{table_widthsu4} how $am_k(L;L_{\perp})$
varies with $L_{\perp}$ for $k=1$ and $k=2$ flux tubes;
all at $\beta=28$ in the SU(4) gauge theory in D=2+1. We do so for 
$L_{\perp}=2,...,20$ and for two values of the flux tube length,
$L$. The minimum transverse size is  
$l_{\perp} = aL_{\perp} = 2a \simeq 0.5/\surd\sigma$ which we 
expect to be smaller than $l_w$. The loop lengths are
$l = aL = 8a,12a \simeq 0.9, 1.35 \mathrm{fm}$ which should be long
enough to allow well-formed flux tubes. The fact that we 
have calculated 
masses for two values of $L$ at each $L_{\perp}$ allows us
to check whether the mass is growing nearly linearly with
$L$ and that we do indeed have a flux tube. We see from the
masses listed in Table~\ref{table_widthsu4} that this is so
for all values of $L_{\perp}$.

We plot the $L=8$ flux loop masses in Fig.\ref{fig_widthsu4d3}.
We see that both $m_{k=1}$ and $m_{k=2}$ start increasing
at a very similar value of $L_{\perp}$. Moreover the
masses for the smallest two values of $L_{\perp}$ are 
consistent with  eqn(\ref{eqn_mwidth}). If we take these 
to fix the value of $L_w$ in eqn(\ref{eqn_mwidth}), we find 
that  $L_w \simeq 5.2$ for both $k=1$ and $k=2$ flux tubes.
That is to say, the flux tube width is indeed independent
of the flux, and has a value $l_w \simeq 1.3/\surd\sigma$.

In addition to these gross features we also see in 
Table~\ref{table_widthsu4} and  Fig.\ref{fig_widthsu4d3}
that there is a decrease in the mass at values of $L_{\perp}$
that are just above the values where the mass starts to
increase. Moreover this decrease is more pronounced for
the $k=2$ string than for the $k=1$ string. Such an effect
indicates that there are some difference between the sizes
of the two flux tubes -- if only in their tails -- and
an analysis of this might provide information on the dynamics, 
e.g. the parameters of the effective dual superconductor
referred to earlier. However anything quantitative needs
calculations with more resolution, i.e at smaller values of $a$.

We display in Fig.\ref{fig_rwidthsu4d3} how $\sigma_{k=2}/\sigma$
varies with $L_{\perp}$. We see that the ratio is close
to Casimir scaling not only at large  $L_{\perp}$ (something
we have seen already) but also at small $L_{\perp}$. 
There is only a small range of  $L_{\perp}$, coinciding
with the dip in $m_k$, where the ratio drops below this. 
Of course, at very small values of $L_{\perp}$ our D=2+1
system is effectively reduced to a D=1+1 gauge-scalar theory, 
and we recall that the linear confinement of  
pure gauge theories in D=1+1 arises through the Coulomb 
interaction which automatically satisfies Casimir scaling.

Our SU(6) analysis closely parallels our SU(4) analysis except 
for the fact that we now have additional $k=3$ strings.
Our calculations are at $\beta=60$ which, as we see from
Table~\ref{table_scaled3}, has a similar $a$ to that
at $\beta=28$ in SU(4).

Our masses are listed in Table~\ref{table_widthsu6}.
They are for flux loops of length $l=10a$ and $l=12a$
and we see evidence for a linearly growing
mass for all $k$ and  for all $L_{\perp}$. We plot the $L=10$
flux loop masses in Fig.\ref{fig_widthsu6d3} and we see,
once again, an increasing loop mass at small $L_{\perp}$
that is consistent with eqn(\ref{eqn_mwidth}). Indeed
we find a common flux tube width, $L_w \simeq 4.5$,
for all three values of $k$. So, just as in SU(4),
the flux tube width is independent of the flux, and has
a value $l_w \simeq 4.5a \simeq 1.2/\surd\sigma$
that is also very similar.
Again, just as in SU(4), the loop mass decreases just
before it begins to increase.

We plot $\sigma_{k}/\sigma$ in  Fig.\ref{fig_rwidthsu6d3}.
We again see consistency with Casimir scaling at small
as well as at large $L_{\perp}$, except possibly in the
region of the dip.

This analysis thus appears to confirm that the confining
flux tube has a cross-section that is largely independent 
of the flux carried. The minor differences between $k$-strings
might, however, be useful in telling us about the details
of the confinement mechanism.

\subsection{High T spatial string tensions}
\label{subsec_highT}

In our above calculations we have calculated the mass
of a long flux loop in a spatial volume with a limited transverse
spatial dimension, $L_{\perp}$. Let us relabel the axes of
our $L\times L_{\perp} \times L_t$ lattice so that the short
spatial torus becomes our time torus and our time torus becomes
a long spatial torus. We are now on a $L_x \times  L_y  \times L_t$
lattice with $L_t = L_{\perp}$ and  $L_x,L_y \gg L_t$.
This corresponds to a system at temperature $aT = 1/L_{\perp}$.
and, as we decrease $L_{\perp}$, we will pass through the deconfining
phase transition at $T=T_c$. In this rotated co-ordinate system
the flux loop `mass' that we have calculated is obtained from the 
spatial correlation of spatial loops; it is a screening mass,
from which we can calculate what is usually referred to as the
`spatial string tension' (usually obtained from spatial Wilson 
loops). Thus our finite width studies have in fact provided 
us with a calculation of the $k=1$, $k=2$ and $k=3$ spatial string 
tensions as a function of $T$ in SU(4) and SU(6) D=2+1 gauge
theories. (All this parallels previous studies 
\cite{MTwidth,MTrot}
of SU(2) in D=2+1.)

Simple arguments (see for example
\cite{MTrot})
tell us to expect $\sigma \propto g^2 T$ for $T\gg T_c$. In our 
case, where $aT=1/L_t= 1/L_{\perp}$, this translates to 
$a^2\sigma \propto 1/L_{\perp}$ as $L_{\perp} \to 0$.
This is precisely what we have already inferred from the
SU(4) and SU(6) calculations in Tables~\ref{table_widthsu4} and
\ref{table_widthsu6}. Indeed we see that the linear 
increase with $T$ sets in very close to $T=T_c$, and does so
simultaneously for all $k$-strings -- presumably
due to the squeezing of a flux tube whose width is $\sim 1/T_c$.
(Although we do not have precise calculations of the deconfining
temperatures in SU(4) and SU(6) gauge theories, we expect
from extrapolations of previous SU(2) and SU(3) calculations
that $T_c \simeq 0.95 \surd\sigma$ in D=2+1 and this tells
us that the critical value of $L_{\perp}$ is $\sim 4$
at these values of $\beta$.)
Moreover, as we see from Figs.~\ref{fig_rwidthsu4d3} 
and~\ref{fig_rwidthsu6d3}, the string tension ratio is close 
to the value expected from Casimir scaling. This is not 
a great surprise: the high-$T$ dimensional reduction of 
the D=2+1 theory takes us to a D=1+1 theory, and in a
D=1+1 gauge theory even the Coulomb potential is linearly
confining; and the latter will automatically satisfy Casimir
scaling. This is of course simplistic; the dimensional
reduction leads to (adjoint) scalars as well as gauge fields, 
in the reduced theory and, in any case, the linear potential
may have other sources than just the Coulomb potential.

We now turn to the more interesting  case of D=3+1. We shall
not attempt a systematic study but, just as in D=2+1,
we shall work at one single value of $\beta$ and will
vary $T$ in the rather coarse steps allowed by varying
$L_t$. Our calculation is in SU(4) at $\beta=10.7$.
Although we do not have precise information on the
deconfining temperature, an extrapolation of previous SU(2) 
and SU(3) calculations
\cite{MTrev}
using a simple $O(1/N^2)$ correction, suggests that
$T_c \simeq 0.62 \surd\sigma$ for SU(4). Since
$a\surd\sigma \simeq 0.306$ at $\beta=10.7$
(see Table~\ref{table_scaled4}) the critical value
of $L_t$ is $\sim 5$. Thus a spatial volume of $10^3$
should be adequately large for our exploratory calculation. 
Accordingly we work on the $10^3 L_t$ lattices listed in 
Table~\ref{table_tempd4a}. In the Table we also include
our earlier calculations on a $10^4$ lattice, to provide 
the `$T=0$' reference value. 

We first wish to establish the rough location of $T_c$. 
To do so we calculate  $\langle l_p \rangle$, the average of 
the thermal line (an unblocked Polyakov loop that winds once 
around the Euclidean time-torus), the lightest mass, $am_t$,
that contributes to spatially separated correlations of
such lines, and $\langle Q^2 \rangle$, the average value of 
the fluctuations of the topological charge $Q$. It is clear
from Table~\ref{table_tempd4a} that there is 
indeed a phase transition close to 
$L_t=5$: the thermal line develops a non-zero vacuum
expectation value, and consequently the lightest mass
becomes $\sim 0$ (the energy of the vacuum). We also see
a very striking suppression of the topological susceptibility,
$a^4\chi_t \equiv \langle Q^2 \rangle/(L^3 L_t)$ across
$L_t=5$. One observes a similar but much less dramatic
behaviour in SU(3)
\cite{MTQsu3},
and this suggests that $\chi_t$ might be an order parameter
for the deconfining phase transition, at least as
$N\to \infty$, with appropriate consequences for the
axial U(1) anomaly. This is a topic we shall expand upon
elsewhere.

Having established what is `high' $T$ in our calculation,
we show in Table~\ref{table_tempd4b} the (screening) 
masses one obtains from spatially separated correlators 
of $k=1$ and $k=2$ loops that wind once around the spatial
torus. At high $T$ we expect $\sigma \propto T^2$ 
(since $g^2$ is now dimensionless). We see that while
our flux loop masses grow faster than $T$ they grow less
fast than $T^2$. This should be no surprise; rather one
should be surprised by the precociously early onset of the
linear high-$T$ behaviour in the case of D=2+1.

From the flux loop masses we calculate the string tension
in two ways: assuming no string correction, i.e. setting
$c_s=0$ in eqn(\ref{eqn_poly}), and assuming a bosonic string
correction. One might expect that at high $T$ one should
use a string correction that is half-way between, because
one has lost one of the transverse dimensions. In any
case it is clear that the high-$T$ ratio is consistent
with Casimir scaling but probably not with the MQCD
formula. Just as for D=2+1 we see a dip in the masses at
$T$ just below $T_c$. By contrast, for this lattice 
calculation, the low-$T$ calculation on the $10^4$ lattice is 
consistent with MQCD but not really with Casimir scaling.  

Once again one might try to use dimensional reduction to
relate the Casimir scaling we observe at high $T$ in D=3+1 
to our observation of it, earlier on in this paper, at low 
$T$ in D=2+1. However any such argument must address the
caveats created by the presence of extra adjoint scalars  
after the reduction. We also note that Casimir scaling
-- at least for $k$-strings and at very high $T$ -- 
has been predicted in a recent model calculation
\cite{ckavor}
which speculates that at high $T$ there is a plasma of
adjoint magnetic pseudoparticles `dual' to the gluon
plasma. (Again one might try to relate
\cite{ckavor}
this to low-$T$ near-Casimir scaling in D=2+1 via 
dimensional reduction.)

\subsection{Unstable strings}
\label{subsec_unstable}

Our calculations have so far focussed on stable $k$-strings. 
In addition to these, there are also unstable strings, which are energetically
unfavourable. An unstable string should appear as a nearly stable excited state
in the $k$-string spectrum. It is clearly interesting to find out how the
string tensions of these strings depend on their representations and on $N$.
We note that this is precisely the question addressed by the
calculations in, for example,~\cite{oldcs,deldar1,bali}. 
In this Section we will investigate closed strings with the quantum numbers
of strings connecting $k$ quarks in a given irreducible representation
for $k=2,3$.

The tensions of strings connecting sources in a given irreducible
representation can be extracted from correlation functions of operators
carrying the quantum numbers of that representation.
For $k=2$ the irreducible representations with two quarks are
the symmetric and the antisymmetric representations.
At $k=3$ the irreducible representations with three quarks
are the totally antisymmetric, the totally
symmetric and the mixed symmetry representation, which enters twice
the decomposition of the tensor product.
The general procedure to construct the relevant operators
for a given representation and the explicit form of the operators
corresponding to the cases we shall investigate in this Section are given in
Appendix~\ref{a1}.

At each ${\cal N}$-ality (and at finite $N$) only the string with the smallest
tension is stable. For any $k$, this string is expected to be the
string connecting $k$ sources in the totally antisymmetric representation.
Hence the smallest mass in the antisymmetric channel is related to the string
tension at the given ${\cal N}$-ality.
Our calculations fulfill this expectation: the smallest
mass extracted with the variational procedure and the smallest mass in the
antisymmetric channel always agree well within errors, both in D=2+1
and D=3+1.

At fixed length an unstable $k$-string is more massive than the stable string
of the same ${\cal N}$-ality. This is likely to give problems
when looking at the exponential decay in time of correlation functions:
the signal will decay too rapidly to allow a reliable extraction of masses.
Indeed this is what happens in our D=3+1 calculations: given the precision
of our numerical data, it proves to be impossible to extract reliable masses.
To overcome this problem, we should get closer to the continuum limit
or use anisotropic lattices. We leave such a study for the future.
Another crucial point is the overlap between the operators and the interesting
states, which if the operators are constructed using standard techniques
(as we have done in the present study) can get as bad as 0.5 for unstable
strings in D=3+1. (For comparison, still in D=3+1, the overlap between the
stringy state corresponding to the antisymmetric representation and our
operators is typically around 0.6.)
Hence, to address questions connected to
unstable strings a better overlap is required.
This requires in turn an improvement of the standard smearing
techniques. This is another problem we will investigate in the future.

While we can not deal at the moment with unstable strings in D=3+1,
our numerical results in D=2+1 allow us to address the question there.
In fact, our D=2+1 calculations do not suffer from the same drawbacks as the
D=3+1 ones: our D=2+1 results are accurate enough to see a clear
exponential decay of the correlation functions over several lattice spacings
and in D=2+1 the overlap between the operators and the unstable strings is not
worse than 0.85. Our numerical results for SU(4) and SU(6) are reported
respectively in Tables~\ref{table_massrepsu4} and~\ref{table_massrepsu6}.
Apart from the symmetric representation
of SU(6), for which we have masses for just two values of the lattice
spacing, we can extrapolate the string tensions extracted from the masses
listed in the Tables to the continuum limit by applying the same procedure
used for stable strings. We then find
\be
\lim_{a \to 0}
{{\sigma_{\mathit{k=2S}}} \over {\sigma}}
=
\left\{ \begin{array}{ll}
2.14 \pm 0.03  & \ \ \ {\mathrm{SU(4)}} \\
2.19 \pm 0.02  & \ \ \ {\mathrm{SU(6)}}
\end{array}
\right. 
\label{eqn_contk2sd3}
\ee
for the string tensions in the $k=2$ symmetric channels and 
\be
\lim_{a \to 0}
{{\sigma_{\mathit{k=3M}}} \over {\sigma}}
=
2.71  \pm 0.09   \ \ \ {\mathrm{SU(6)}}
\label{eqn_contk3md3}
\ee
for the $k=3$ mixed symmetry channel. (All these 
string tensions have been extracted
using the bosonic string correction.) For the continuum value
of the $k=3S$ string tension, an estimate based on our data gives
\be
\lim_{a \to 0}
{{\sigma_{\mathit{k=3S}}} \over {\sigma}}
\approx
3.72  \pm 0.12     \ \ \ {\mathrm{SU(6)}} .
\label{eqn_contk3sd3}
\ee
Our numerical results can be compared with the predictions coming from
Casimir scaling
\be
{{\sigma_{\mathit{k=2S}}} \over {\sigma}}
\stackrel{CS}{=}
\left\{ \begin{array}{ll}
2.40 & \ \ \ {\mathrm{SU(4)}} \\
2.28...  & \ \ \ {\mathrm{SU(6)}}
\end{array}
\right. 
\!\! ,
\label{eqn_cask2sd3}
\ee
\be
{{\sigma_{\mathit{k=3M}}} \over {\sigma}}
\stackrel{CS}{=}
2.82...  \ \ \ {\mathrm{SU(6)}} ,
\label{eqn_cask3md3}
\ee
\be
{{\sigma_{\mathit{k=3S}}} \over {\sigma}}
\stackrel{CS}{=}
3.85...   \ \ \  {\mathrm{SU(6)}} . 
\label{eqn_cas3sd3}
\ee
We see that, at least for SU(6), these ratios satisfy approximate Casimir
scaling, just like the stable $k$-strings. As far as comparison with
MQCD is concerned, we are not aware of
calculations in that framework aimed to determine the string tensions of
unstable strings.
\section{Discussion}
\label{sec_discussion}

Our calculations in this paper were in two parts. 
In the first part we investigated directly the 
stringy nature of long flux tubes by calculating
how the mass of a flux tube varies with its length $l$,
and attempting to identify the $O(1/l)$ term that
is the leading string correction at large $l$.
The coefficient of this term, $c_L$, is directly related
to the central charge of the effective string
theory describing the long-distance physics of
the confining flux tube, and thus characterises its
universality class. By considering flux tubes that
wind around a spatial torus we were able to 
avoid the presence of explicit sources and the
accompanying Coulomb term which can be so easily 
confused with the string correction. Working in
SU(2) and at a reasonably small value of the lattice
spacing $a$, we obtained in 3+1 dimensions a value
$c_L=0.98\pm0.04$ which is consistent with
the simple bosonic string, for which $c_L =\pi/3$. In 2+1 
dimensions we obtained $c_L=0.558 \pm 0.019$, which is again 
consistent with the bosonic string value, which is
$c_L =\pi/6$ in this case. In both dimensions our results 
would appear to exclude other plausible possibilities with, 
for example, some massless fermionic modes along the string.
In addition, in the case of D=2+1 our results were accurate 
enough to constrain the power of $1/l$ to be unity (assuming
it to be an integer) as one expects for an effective string
theory. These results  considerably increase, we believe, 
the evidence for the simple bosonic string model.
There is however much scope for improving these calculations;
not only in their accuracy and in the range of flux tube lengths
studied, but also in exploring other values of $a$, so as to be
confident of the continuum physics, and in extending the
calculations to other SU($N$) groups.

The second part of the paper dealt with $k$-strings in
SU($N\geq4)$ gauge theories and, in particular, with the
ratios of their string tensions, $\sigma_k/\sigma$.
Here we performed a range of calculations so as to be 
able to extrapolate to the continuum limit. In our D=3+1
SU(4) and SU(5) calculations we found that the 
$k=2$ string tension is much less than twice the 
fundamental $k=1$ tension: the $k$-string is `strongly bound'.
Moreover the values are consistent, at the $2\sigma$ level, 
with both the M(-theory)QCD 
conjecture and with Casimir scaling (the two being numerically
quite similar). In our SU(4) and SU(6) calculations in D=2+1 
we also found strongly bound $k$-strings. However, although
the calculated string tension ratios were again numerically
close to both Casimir scaling and the MQCD formula, the results 
were accurate enough for us to see that the former 
works much better, and to observe deviations from both
formulae. In addition to these continuum calculations we
performed some finite temperature calculations at fixed $a$
which showed that at high $T$, above the deconfinement 
transition, `spatial' $k$-string tensions are consistent with 
Casimir scaling in both 2+1 and 3+1 dimensions. Moreover we found
fairly convincing evidence, in D=2+1, for the approximate Casimir 
scaling of unstable strings. While it might be elegant
if (approximate) Casimir scaling were to hold in D=3+1 
as well as in D=2+1,
and at high $T$ as well as at low $T$, the fact is that
3+1 dimensions may well differ from 2+1 dimensions, and
it is important to perform calculations that are accurate
enough to resolve between MQCD and Casimir scaling in
D=3+1. Essentially this would require reducing our errors
by a factor of two, an entirely feasible goal. 

We observed that near-Casimir scaling will arise naturally
if the chromo-electric flux is homogeneous and the cross-section 
of the flux tube is (nearly) independent of the flux carried.
We pointed out that the latter is not as implausible as it might
at first seem: indeed it is what occurs in the deep-London 
limit of a superconductor. To address this possibility
we performed some explicit numerical calculations of
the $k$-string width and these indicated that the
width is indeed largely independent of $k$. The small $k$ 
dependence that we did observe can, in principle, be related to
the parameters of the dual superconductor, if such is the
dynamics of confinement, and we intend to
address this question elsewhere. There are a number of other
interesting theoretical questions that this work suggests.
How closely is the observed near-Casimir scaling of the
(`spatial') $k$-strings at high-$T$ in D=3+1 and 
at low $T$ in D=2+1 related by dimensional reduction? 
Equally, is the near-Casimir scaling at high-$T$ 
in D=2+1, a simple reflection of the Casimir scaling
of the linearly confining Coulomb interaction in D=1+1?
This requires understanding whether the adjoint scalars,
present after dimensional reduction, significantly
affect the string tension ratios. Another interesting
question is how the string tension ratios, whether
given by MQCD or Casimir scaling, reflect themselves
in $k$-vortex condensates, in the dual disorder
loop approach to confinement
\cite{thooftvor},
and whether this exposes any simple duality between Wilson
loops and 't Hooft disorder loops. A calculation, illustrating 
how one might proceed, was outlined in 
\cite{string00,MTrot}.
A similar question can be posed in monopole models of 
confinement, following upon the simple model calculations
of higher charged string tensions in
\cite{ahmtmon},
after Abelian projection, and in
\cite{ckavor}
with adjoint monopoles (at high $T$). A quite different
question is what are the implications of these 
tightly-bound $k$-strings for the mass spectrum of
SU($N$) gauge theories. A simple and attractive model
sees the glueball spectrum as arising from excitations
of closed loops of fundamental flux
\cite{ip}.
In such a model a non-trivial $k$-string would provide
a new sector of states whose masses are scaled up
by a simple factor of $\sigma_k/\sigma$ 
\cite{mtrjip}.
The observation of something like this, when comparing the 
SU(3) and SU(4) spectra for example, would provide striking 
information on glueball structure. While our D=3+1 
mass spectrum calculations
\cite{glue01}
are too crude to usefully explore this question, this is
not the case in D=2+1 (see e.g.
\cite{mtd3})
and work on this question is proceeding.

\vspace{1.5cm}
{\underline{Note added:}} As this paper was being completed,
a paper appeared 
\cite{pisa01}
containing a calculation of $k=1,2,3$ string tensions in
D=3+1 SU(6) and addressing some of the questions addressed in 
this work.

%
%
\section*{Acknowledgments}
We are grateful to Gunnar Bali, Luigi Del Debbio, Ian Kogan, Ken Konishi,
Chris Korthals Altes, Alex Kovner, Simone Lelli,
Paolo Rossi, Riccardo Sturani, Bayram Tekin, Alessandro Tomasiello,
Ettore Vicari and numerous other colleagues,
for useful communications and discussions. Our calculations
were carried out on Alpha Compaq workstations in Oxford
Theoretical Physics, funded by PPARC and EPSRC grants.
One of us (BL) thanks PPARC for a postdoctoral fellowship.


\vspace{1.5cm}

\appendix
\section*{Appendix}
\addtocounter{section}{1}

This Appendix collects detailed proofs of some statements contained in the
main exposition. In Section~\ref{a1} we will derive the explicit form
of the operators carrying the quantum numbers of $k$-strings.
In Section~\ref{a3} we will show how sources in a given representation can
be screened by gluons. Finally, Section~\ref{a2}
will deal with the quadratic Casimirs of irreducible representations of
SU($N$) and their relationship with Casimir scaling.
Our calculations will be heavily based on Group Theory.
In order to make this Appendix self-contained, we will recall some general
results of Group Theory. For a wider introduction to the group theoretical
background we refer to~\cite{grouptheory}.

\subsection{Irreducible representations of SU($N$) and k-stings}
\label{a1}
SU($N$) is the group of $N \times N$ unitary matrices. By definition, an object
$q^i \ (i=1,\dots, N)$ transforms under the fundamental representation of
SU($N$) if under the action of the group
\be
q^i \mathop{\to}^{SU(N)} U_j^i q^j , 
\ee
$U \in$ SU($N$) being the matrix that implements the transformation.

The conjugated representation is related to the fundamental one by complex 
conjugation. Following the standard notation, we indicate by $q_i$ an object 
transforming under the conjugated representation. For $N \ge 3$ the fundamental
and the conjugated representations are independent.
In the following, we will call {\em quarks} objects transforming
under the fundamental representation and {\em antiquarks} objects
transforming under the conjugated representation of SU($N$).
This terminology reflects the physics of QCD.

Objects transforming under higher representations can be constructed from the
tensor product of quarks and antiquarks, and their transformation laws can
be easily deduced from the transformation law of the fundamental constituents.
For instance $(q \otimes q)^{ij} = q^i q^j$ under the action of SU($N$)
transforms as follows:
\be
q^i q^j \mathop{\to}^{SU(N)} U_k^i U_l^j q^k q^l .
\ee

The ${\cal N}$-ality of a representation is defined as the number of quarks
minus the number of antiquarks modulo $N$. ${\cal N}$-alities
$k \le N/2$ and $N-k$ are related
by complex conjugation. That operation corresponds to charge conjugation.

The concept of ${\cal N}$-ality at zero temperature is related to the
symmetry under the centre of the gauge group, $Z_N$:
under such symmetry an object of ${\cal N}$-ality $k$ pick up a phase
$e^{(2\pi i k n)/N}$, $n=0,1,\dots,N-1$. Since the centre symmetry at zero
temperature is a good symmetry of the gauge theory and the gluons carry zero
${\cal N}$-ality, states with different ${\cal N}$-ality cannot
mix.

In this paper, we are interested in the tensions of strings connecting
sources with ${\cal N}$-ality $k \ge 1$. Because of charge conjugation,
the string tension associated to
states of ${\cal N}$-ality $k \le N/2$ and $N-k$ is the same. Hence we
restrict ourself to $k \le N/2$, that is to say to states constructed from
the tensor product of $k$ quarks\footnote{Here we are neglecting states with
more than $N$ quarks; this is correct as far as we are not interested in
unstable strings or $N$ is large enough.}. At a given $N$ the independent
number of stable $k$-strings is given by the integer part of $N/2$.

As for the fundamental string, the string tension of a $k$-string can be
extracted by looking at the exponential decay of correlators of loop operators
with the quantum numbers of that string. In order to identify the
relevant operators, it is useful to decompose the tensor product of $k$ quarks
into irreducible representations\footnote{In the following, even if we will
omit for simplicity the word {\em irreducible} from time to time,
we will consider only irreducible representations of the gauge group.}.
To this purpose, the Young tableau technique can be used.

A Young diagram is a two-dimensional ensemble of boxes joined by one edge
that respects the following rules:
\begin{enumerate}
\item counting the rows from the top to the bottom, the number of boxes in the
row $i$ is greater than or equal to the number of boxes in the row
$j$ if $i<j$;
\item counting the columns from the left to the right, the number of boxes in
the column $i$ is greater than or equal to the number of boxes in
the column $j$ if $i<j$.
\end{enumerate}
A valid Young tableau is for instance the following:\\
\begin{center}
\epsfig{figure=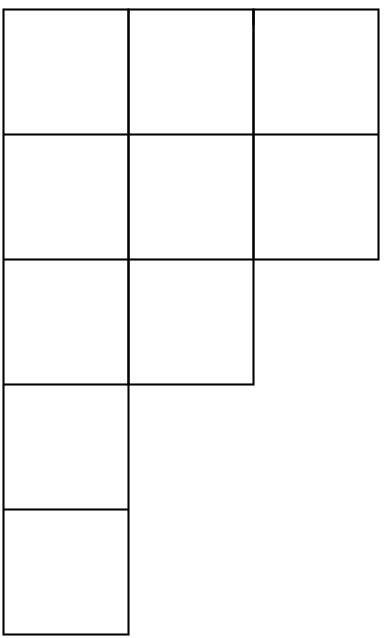, height=3cm} 
\end{center}
In the Young tableau language, a quark is a single box, an antiquark is
a column of $N-1$ boxes and an object transforming under the adjoint
representation ({\em gluon}) has $N-1$ boxes in the first column and 
2 boxes in the first row:
\begin{center}
\begin{tabular}{ccccc}
\epsfig{figure=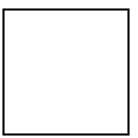, height=0.6cm} & \ \ \ &
\epsfig{figure=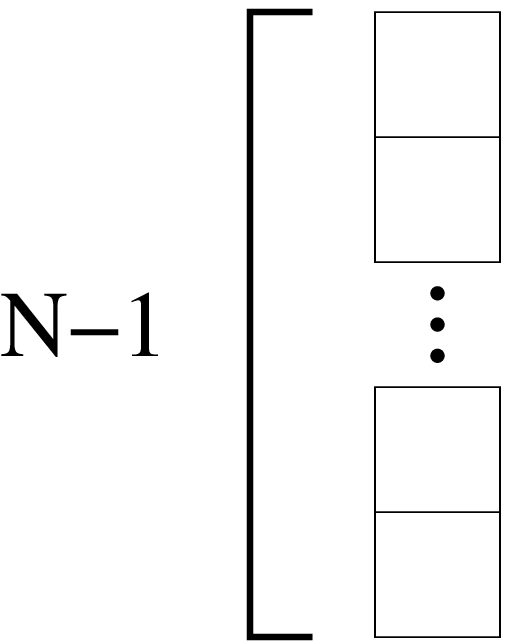, height=3cm} \ \ \ &
\epsfig{figure=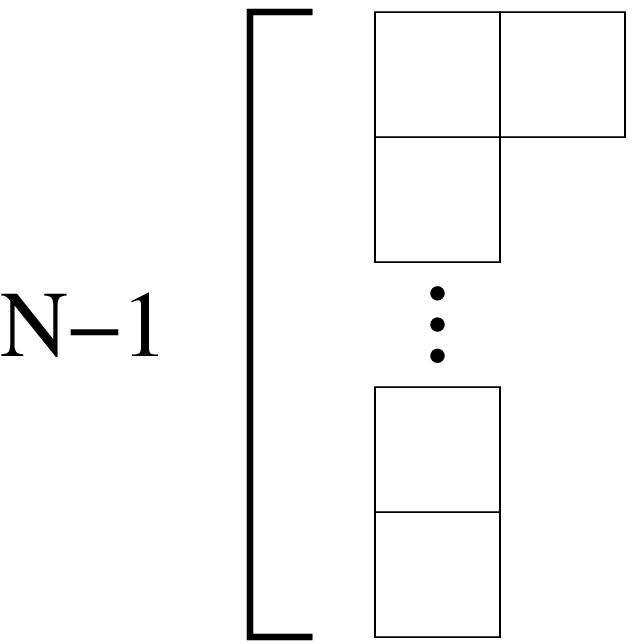, height=3cm} \\
{\em quark} & \ \ \ & \ \ \ \ {\em antiquark} & & \hspace{-2cm}
{\em gluon}
\end{tabular}
\end{center}

There is a one-to-one correspondence between Young diagrams and irreducible
representations of SU($N$). Given a Young diagram with $k$ boxes, the object
transforming under the corresponding irreducible representation is
constructed from the tensor product of $k$ quarks by assigning an index to
each box, symmetrising the
product with respect to the indices that are on a given row for all rows and
then antisymmetrising the result with respect to the indices that are on a
given column for all columns. (Obviously, after the antisymmetrisation the
result is no longer symmetric under permutation of indices on the same row.)

The tensor product of two objects transforming under two given representations
of the gauge group is constructed from the corresponding Young diagrams,
$A$ and $B$, according to the following rules:
\begin{enumerate}
\item write down the two tableaux $A$ and $B$ labelling each box
in the row $i$ of $B$ by $i$;
\item starting from the first row of $B$,
add the boxes of $B$ to $A$ one-by-one in all the possible positions
respecting the following rules: 
\begin{enumerate}
\item the augmented diagram $A^{\prime}$ at each stage must be a legal Young
diagram;
\item boxes with the same label must not appear in the same column of
$A^{\prime}$;
\item If we define at any given box position $J$ numbers
$n_1,\dots,n_J$ ($J$ being the number of rows in $B$), each of them counting
how many times the corresponding label of the boxes in $B$ appears above
and to the right of such a box, we must have $n_1 \ge n_2 \ge \dots \ge n_J$
(this is to take into account the original antisymmetries of $B$);
\end{enumerate}
\item two diagrams with the same shape and the same labels are the same
diagram;
\item columns with $N$ boxes must be canceled, since they correspond to the
trivial representation of SU($N$).
\end{enumerate}

According to the above rules, the tensor product of two quarks decomposes
as
\begin{center}
\epsfig{figure=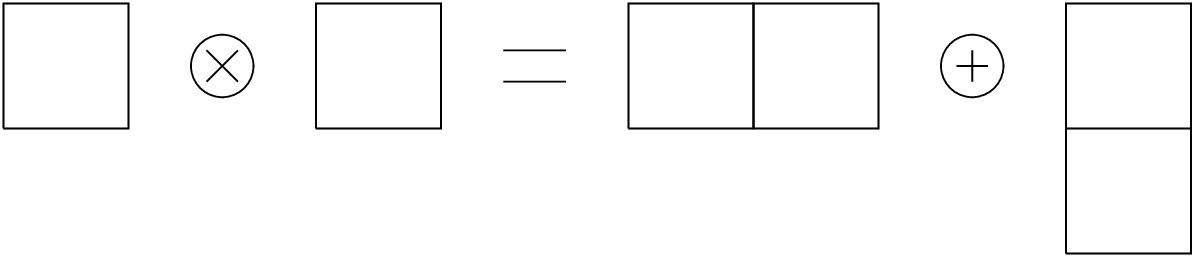, angle=270, width=5.2cm} 
\end{center}
i.e. the irreducible representations of a state with two quarks are the
symmetric and antisymmetric representations. For three quarks we have\\ 
\begin{center}
\epsfig{figure=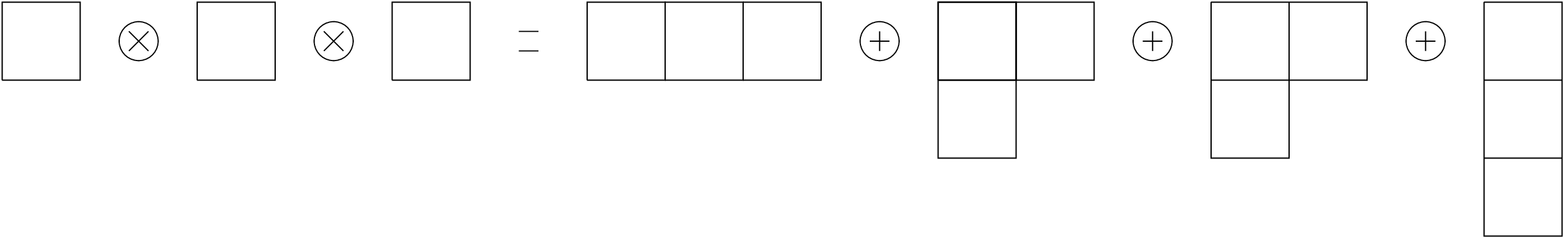, angle=270, width=12cm} 
\end{center}
where in addition to the symmetric and antisymmetric representation
there is a representation with mixed symmetry entering twice the decomposition.

The above results are the generalisation in SU($N$) of the familiar
decompositions in SU(3) $3 \otimes 3 =
6 \oplus \bar{3}$ and $3 \otimes 3 \otimes 3 = 10 \oplus 8 \oplus 8 \oplus 1$.

Once the symmetry of the states transforming under an irreducible
representation has been worked out, it is easy to construct
the operators implementing the transformation on such states, since those
operators must have the same symmetry as the states on which they act.
For the matrix elements of $k=2$ operators associated to strings
connecting sources with two quarks we obtain
\begin{eqnarray}
A^{ij}_{lm} &=& \frac{1}{2} \left(U^{i}_{l} U^{j}_{m} - U^{j}_{l} U^{i}_{m}
\right) , \\
S^{ij}_{lm} &=& \frac{1}{2} \left(U^{i}_{l} U^{j}_{m} + U^{j}_{l} U^{i}_{m}
\right) ,
\end{eqnarray}
while for $k=3$ strings connecting three quarks we have
\begin{eqnarray}
A^{ijk}_{lmn} &=& \frac{1}{6} \left(U^{i}_{l} U^{j}_{m} U^{k}_{n} - 
U^{i}_{l} U^{k}_{m} U^{j}_{n} - U^{j}_{l} U^{i}_{m} U^{k}_{n}
+ U^{j}_{l} U^{k}_{m} U^{i}_{n} + U^{k}_{l} U^{i}_{m} U^{j}_{n} -
U^{k}_{l} U^{j}_{m} U^{i}_{n} \right) , ~~~~~~\\
S^{ijk}_{lmn} &=& \frac{1}{6} \left(U^{i}_{l} U^{j}_{m} U^{k}_{n} + 
U^{i}_{l} U^{k}_{m} U^{j}_{n} + U^{j}_{l} U^{i}_{m} U^{k}_{n}
+ U^{j}_{l} U^{k}_{m} U^{i}_{n} + U^{k}_{l} U^{i}_{m} U^{j}_{n} +
U^{k}_{l} U^{j}_{m} U^{i}_{n} \right) , \\
\label{mixed}
M^{ijk}_{lmn} &=& \frac{1}{3} \left(U^{i}_{l} U^{j}_{m} U^{k}_{n}
- U^{k}_{l} U^{j}_{m} U^{i}_{n} + U^{j}_{l} U^{i}_{m} U^{k}_{n}
- U^{j}_{l} U^{k}_{m} U^{i}_{n}  \right) ,
\end{eqnarray}
$A$, $S$ and $M$ being respectively the tensors corresponding to the
antisymmetric, the symmetric and one of the two mixed symmetry
representations.~(The other mixed symmetry representation has $k$ and $j$
interchanged in eqn(\ref{mixed}).)

Taking the trace, we get
\begin{eqnarray}
Tr A &=& \frac{1}{2} \left( \{Tr U\}^2 - Tr U^2 \right) , \\
Tr S &=& \frac{1}{2} \left( \{Tr U\}^2 + Tr U^2 \right) ,
\end{eqnarray}
for k=2 and 
\begin{eqnarray}
Tr A &=& \frac{1}{6} \left( \{Tr U\}^3 - 3 Tr U\{Tr U\}^2
+ 2 Tr U^3 \right) , \\
Tr S &=& \frac{1}{6} \left( \{Tr U\}^3 + 3 Tr U\{Tr U\}^2
+ 2 Tr U^3 \right) , \\
Tr M &=& \frac{1}{3} \left( \{Tr U\}^3 - Tr U^3 \right) ,
\end{eqnarray}
for k=3. (The two different $M$'s have the same trace.)

After identifying $U$ with the path ordered product of links around
a non-contractible loop $c$ that winds once around the spatial
torus, $P_c$, we get that the relevant operators for $k=2$ are
$Tr P^2_c$ and $\{Tr P_c\}^2$, while for $k=3$ we will be concerned with
$Tr P^3_c$, $Tr P_c\{Tr P_c\}^2$ and $\{Tr P_c\}^3$. These operators
can be taken as a starting point for a variational procedure to extract
the mass of flux tubes of ${\cal N}$-ality $k$ winding once around the
periodic lattice, while studying directly the combination corresponding
to a given irreducible
representation is relevant in the context of unstable strings
(see the following Section).

The construction here explicitly provided for $k=2$ and $k=3$ can be easily
generalised to any $k$.
\subsection{Gluon screening}
\label{a3}
Since the gluons transform under a non-trivial representation of SU($N$),
the interaction between them and the sources can change the original
representation of the sources. The change of representation of the source
is expected to renormalise the string tension associated to the original
representation and to make unstable heavier strings of given ${\cal N}$-ality. 

From the point of view of Group Theory, the product of the interaction between
sources and gluons transforms as the tensor product of the original
representation and the adjoint representation. Consider for instance the
following interaction:\\
\begin{center}
\epsfig{figure=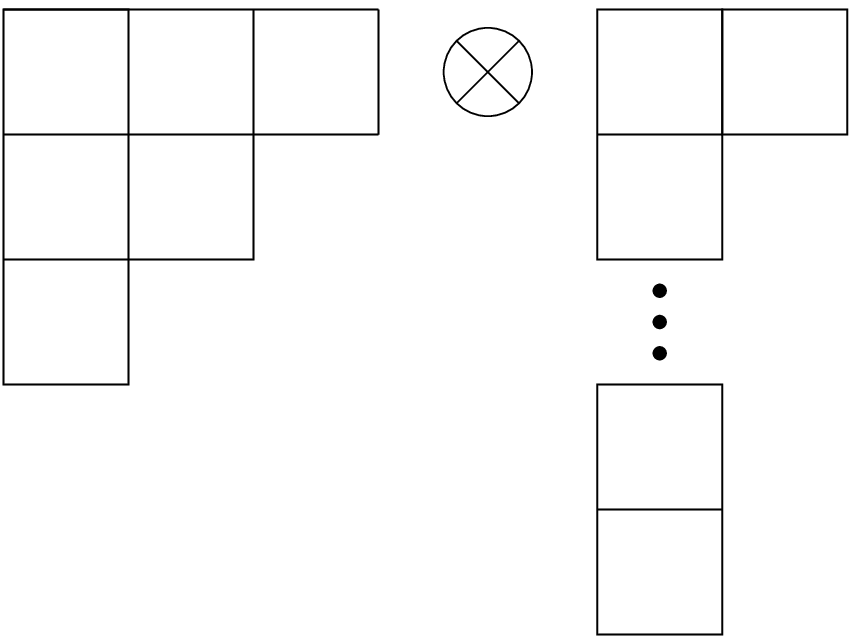,height=3cm}
\end{center}
where the first tableau correspond to a source with $k$ quarks and the second
diagram is associated to a gluon. The interaction  will produce an object
transforming under a reducible representation. The irreducible representations
entering the product of the interaction can
be worked out according to the rules for the decomposition of a tensor product
given in the previous section. Those representations fall into two categories:
representations with $k$ quarks and representations with $N+k$ quarks. 

Let us consider the first case. In order to have a final state with $k$ quarks,
in the tensor product $N$ boxes must be canceled (i.e. they are combined in
such a way that they transform under the trivial representation). Given the
diagram of a source with $m$ boxes in the first column, the
cancellation of $N$ boxes in the tensor product requires that $N-m$ boxes
from the gluon are attached to the first column of the source. This can be done
in two inequivalent ways: by taking the required objects all from rows other
than the first one or by taking one box from the first row\footnote{This
argument should be refined if we were interested to the multiplicity
with which each irreducible representation enters the decomposition of the
tensor product.}. The possible ways
of recombining the diagrams after the cancellation define possible
representations of the interacting state. Those representations depend on
the original representation of the sources, but not on $N$. Similar
considerations hold for the irreducible representations with $N+k$ quarks
entering the decomposition.

An interacting state will be energetically favourable whenever it has a smaller
string tension, so we expect that the interaction tends to transform
the sources in a given representation of ${\cal N}$-ality $k$ to sources in the
representation with the smallest string tension (which is the antisymmetric
representation with $k$ quarks in both the Casimir scaling and the MQCD
scenario), i.e. that the gluons screen the sources down to the states with the
smallest string tension\footnote{Note however that there is a phase-space
suppression factor for the screening of sources with a given number of
quarks down to sources with a minor number of quarks inside the same
${\cal N}$-ality class. Consider for instance the screening of a representation
with $N+k$ quarks to a representation with $k$ quarks. Since the dimension of
the former is proportional to $N^{k+2}$, while the dimension of the latter
is proportional to $N^k$, there is a suppression due to the lack of final
states proportional to $1/N^2$.}. However unstable strings are expected to
be visible, since they should appear as nearly-stable excited states in the
mass spectrum of the strings.

Note that not all states with ${\cal N}$-ality $k$ are accessible to a given
state. For instance, the interaction with one gluon does not allow to pass
from the symmetric to the antisymmetric representation at $k=3$.
However any state can be accessed by multiple interaction.
\subsection{Quadratic Casimir operator and Casimir Scaling}
\label{a2}
The quadratic Casimir operator of a representation ${\cal R}$
is defined as
\be
c_{\cal R} \equiv \sum_a T^a T^a 
\ee
where the sum ranges over all the generators of the group in the given
representation. It can
be easily seen that the quadratic Casimir operator commutes with all the
generators of the group. Hence, by virtue of the Schur's lemma, on a given
representation it is proportional to the identity, i.e. it is identified by a
number depending on the representation. We call such
a number quadratic Casimir and we indicate it by $C_{\cal R}$. If we normalise
the trace of the identity in each representation to 1, we can write 
\be
C_{\cal R} \equiv Tr T^a T^a .
\ee

$C_{\cal R}$ can be easily computed starting from the Young tableau
associated to the representation ${\cal R}$ as follows. For SU($N$), defines
the $N$-dimensional vectors
\begin{eqnarray}
\nonumber
\vec{L}_1 &=& \frac{1}{N} \left(N-1,-1,-1,-1,\dots,-1\right) , \\ 
\nonumber
\vec{L}_2 &=& \frac{1}{N} \left(N-2,N-2,-2,-2,\dots,-2\right) , \\ 
\nonumber
\vec{L}_3 &=& \frac{1}{N} \left(N-3,N-3,N-3,-3,\dots,-3\right) , \\ 
\nonumber
&& \vdots \\ 
\nonumber
\vec{L}_{N-1} &=& \frac{1}{N} \left(1,1,1,,1,\dots,-(N-1)\right) , \\ 
\nonumber
2\vec{R} &=& \left(N-1,N-3,N-5,\dots,-(N-3),-(N-1)\right) .
\end{eqnarray}
With the vectors $\vec{L}_i$, define
\be
\nonumber
\vec{L} = \sum_{i=1}^{N-1} w_i L_i 
\ee
where $w_i$ is given by the difference between the number of boxes in the
row $i$ and the number of boxes in the row $i+1$ of the Young tableau.
The quadratic Casimir is then given by
\be
\nonumber
C_{\cal R} = \frac{1}{2} \left(
\vec{L}\cdot\vec{L} + 2 \vec{R}\cdot\vec{L} \right) .
\ee
It is now easy to see that for an irreducible representation composed by 
$k$ quarks the quadratic Casimir is given by the formula
\be
\label{generalcasimir}
C_{\cal R} = \frac{1}{2} \left( Nk + \sum_{i=1}^{m} n_i (n_i + 1 -2i) -
\frac{k^2}{N} \right) 
\ee
where $i$ ranges over the rows of the Young tableau (with $m$ number of rows)
and $n_i$ is the number of boxes in the $i$-th row.
For the antisymmetric and the symmetric representations of ${\cal N}$-ality $k$
we have
\be
C_A = C_f \frac{k(N-K)}{N-1}
\ee
and
\be
C_S = C_f \frac{k(N+K)}{N+1} ,
\ee
\be
C_f = \frac{N^2 -1}{2N}
\ee
being the quadratic Casimir of the fundamental representation.

For $k=3$ in addition to the symmetric and antisymmetric representations, there
is (among others) the mixed symmetry representation, whose quadratic Casimir
is
\be
C_M = C_f \frac{3 (N^2 -3)}{N^2 - 1} .
\ee

Casimir scaling is the hypothesis that the string tension for a given
representation is proportional to the quadratic Casimir. Hence, according to
this hypothesis, at ${\cal N}$-ality $k$ the smallest string tension is
associated to sources in the representation with the smallest quadratic
Casimir. By using
eqn(\ref{generalcasimir}), it can be easily seen that the representation
having the smallest quadratic Casimir is the totally antisymmetric
representation composed by $k$ quarks. To show this, let us prove as a
preliminary step that if we increase the antisymmetries of a diagram
keeping constant the number of boxes the quadratic Casimir decreases. 
In fact the difference between the quadratic Casimir of a given representation
and of the representation obtained by moving a box of the original Young
diagram from the $j$-th row to the $h$-th row with $h>j$ is
\be
\Delta C = n_j - n_h + h  - j - 1  > 0 .
\ee
It is now easy to prove the main statement: at given number of boxes $k \le
N/2$, the antisymmetric representation is obtained from any given
representation by iterating the above procedure, with a series
of steps where at each stage the quadratic Casimir decreases. Thus  
at given $k$ the antisymmetric representation with $k$ quarks
has the smallest quadratic Casimir and for this reason the smallest string
tension within the class of the representations with ${\cal N}$-ality $k$
in the Casimir scaling hypothesis.
This fact does old even if we consider states with $N+k$
quarks: the difference between the quadratic Casimir of the most antisymmetric
representation with $N+k$ quarks (which is the smallest at that number of
quarks) and the totally antisymmetric representation with $k$ quarks is 
\be
\Delta C = N-k .
\ee

Not all the representations with $k$ quarks have a larger quadratic Casimir
than a given representation with $N+k$ quarks. For instance, the difference
between the quadratic Casimir of the most antisymmetric representation
with $N+k$ quarks and the totally symmetric representation with $k$ quarks
is
\be
\Delta C = N-k^2 ,
\ee
which is negative if $k^2>N$. However, at large enough $N$ and at a given
$k$ such a difference is positive and increases as $N$.
That is to say, we expect that the only relevant states of ${\cal N}$-ality
$k$ in the large $N$ limit are those composed by $k$ quarks.

The prediction of Casimir scaling for the ratio between the string tensions
associated to states composed by $N+k$ quarks and the string tension of the
fundamental representation in the
limit $N \to \infty$ is $k+2$. This result can be easily understood in terms
of string counting: a state with $N+k$ quarks can be seen as a possible state
among those originated by the interaction between a state with $k$ quarks
and a gluon.
The above result then tells us that at large $N$ the energy of the composite
state is equal to the sum of the energies of the constituents.

\vfill \eject

\begin{table}
\begin{center}
\begin{tabular}{|c|c||c|c|c|}\hline
\multicolumn{5}{|c|}{ $am_l$ ; SU(2) ; D=2+1} \\ \hline
lattice & MC sweeps & $p=0$ & `safe' $p=0$ & low $p$  \\ \hline
$8^2 64$  & $4 \times 10^5$ & 0.1703(4) & 0.1703(4) & 0.1703(4) \\
$10^2 48$ & $8 \times 10^5$ & 0.2167(5) & 0.2167(5) & 0.2167(5) \\
$12^2 36$ & $10^6$          & 0.2696(5) & 0.2696(5) & 0.2696(5) \\
$14^2 36$ & $2\times 10^6$  & 0.3219(6) & 0.3210(8) & 0.3219(6) \\ 
$16^2 32$ & $2\times 10^6$  & 0.3812(4) & 0.3806(6) & 0.3812(4) \\
$20^2 32$ & $2\times 10^6$  & 0.4917(7) & 0.4906(9) & 0.4917(5) \\
$24^2 32$ & $2\times 10^6$ & 0.5998(13) & 0.5998(13) & 0.6004(8) \\
$28^2 32$ & $2\times 10^6$ & 0.7101(10) & 0.7089(18) & 0.7083(9) \\
$32^2 32$ & $2\times 10^6$ & 0.8131(23) & 0.8131(23) & 0.8150(15) \\
$36^2 32$ & $2\times 10^6$ & 0.9175(33) & 0.9167(38) & 0.9195(19) \\
$40^2 32$ & $2\times 10^6$ & 1.0238(53) & 1.0238(53) & 1.0284(24) \\
\hline
\end{tabular}
\caption{\label{table_d3su2_stringm}
The lightest mass $am_l$ of a fundamental string wrapped 
around a spatial torus. The first column comes from good
fits to $p=0$ correlators, chosen so as to minimise the 
errors. The second column contains cautious `very safe' 
estimates with larger errors. The third column uses
both $p=0$ and, where useful, $p\not= 0$ correlators.}
\end{center}
\end{table}

\begin{table}
\begin{center}
\begin{tabular}{|c|c||c|c|c|}\hline
\multicolumn{5}{|c|}{ $c_s^{eff}(l,l^{\prime})$ ; SU(2) ; D=2+1} \\ \hline
$L$ & $L^{\prime}$ & $p=0$ & `safe' $p=0$ & low $p$  \\ \hline
 8  & 10 & 0.130(24) & 0.130(24) & 0.130(24) \\
 10 & 12 & 0.498(41) & 0.498(41) & 0.498(41) \\
 12 & 14 & 0.546(62) & 0.479(73) & 0.546(62) \\
 14 & 16 & 1.329(79) & 1.37(11)  & 1.329(79) \\
 16 & 20 & 1.046(58) & 1.01(8)   & 1.032(48) \\
 20 & 24 & 0.99(16)  & 1.15(18)  & 1.08(10)  \\
 24 & 28 & 1.53(27)  & 1.35(35)  & 1.16(19) \\
 28 & 32 & 0.31(51)  & 0.6(6)    & 1.10(36) \\
 32 & 36 & 0.7(1.1)  & 0.5(1.2)  & 0.68(66) \\
 36 & 40 & 1.4(2.1)  & 1.7(2.2)  & 2.2(1.0) \\
\hline
\end{tabular}
\caption{\label{table_d3su2_cstring1}
Estimating the effective string correction coefficient from 
the masses of pairs of flux loops of length $l=aL$ and 
$l^{\prime}=aL^{\prime}$
respectively, using eqn(\ref{eqn_polypair}).}
\end{center}
\end{table}

\begin{table}
\begin{center}
\begin{tabular}{|c||cc|cc|cc|}\hline
\multicolumn{7}{|c|}{ $c_s(l\geq l_0)$ ; SU(2) ; D=2+1} \\ \hline
$L_0$ & $p=0$ & $\chi^2/dof$ & `safe' $p=0$ & $\chi^2/dof$ & 
low $p$ & $\chi^2/dof$ \\ \hline
 14 &  --       &  -- & 1.118(28) & 1.3 & 1.104(18) & 1.6 \\
 16 & 1.071(26) & 1.2 & 1.066(36) & 0.4 & 1.070(22) & 0.4 \\
 20 & 1.091(63) & 1.4 & 1.11(8) & 0.5 & 1.105(42)   & 0.3 \\
 24 & 1.02(16) & 1.9 & 1.01(16) & 0.5 & 1.12(9)  & 0.4 \\
 28 & 0.52(28) & 0.2 & 0.69(35) & 0.1 & 1.10(17) & 0.6 \\
 32 & 0.9(7)   & 0.1 & 0.88(75) & 0.2 & 1.21(40) & 1.0 \\
\hline
\end{tabular}
\caption{\label{table_d3su2_cstring2}
Estimating the string correction coefficient from a
fit of eqn(\ref{eqn_poly}) to the masses of all the
flux loops  of length $l\geq l_0 = a L_0$. In each case we
show the $\chi^2/dof$ of the best fit.}
\end{center}
\end{table}

\begin{table}
\begin{center}
\begin{tabular}{|c|c||c|c|c|c|}\hline
\multicolumn{6}{|c|}{ $am_l$ ; SU(2) ; D=3+1} \\ \hline
lattice & MC sweeps & $p=0$ & `safe' $p=0$ & low $p$  
& `safe' low $p$  \\ \hline
$10^3 60$ & $10^5$         & 0.1679(14) & 0.1679(14) & 0.1679(14)
& 0.1679(14) \\
$12^3 48$ & $2\times 10^5$ & 0.2073(14) & 0.2073(14) & 0.2073(14)
& 0.2073(14) \\
$14^3 36$ & $4\times 10^5$ & 0.2632(13) & 0.2606(16) & 0.2636(12)
& 0.2606(16) \\
$16^3 28$ & $6\times 10^5$ & 0.3230(18) & 0.3230(18) & 0.3302(11) 
& 0.3230(18) \\
$20^4$    & $6\times 10^5$ & 0.4468(15) & 0.4416(23) & 0.4408(20) 
& 0.4408(20) \\
$24^4$    & $8\times 10^5$ & 0.5476(22) & 0.5476(22) & 0.5459(15) 
& 0.5459(15) \\
$32^3 24$ & $4\times 10^5$ & 0.7598(75) & 0.7469(115) & 0.7549(50)  
& 0.7496(58) \\ \hline
\end{tabular}
\caption{\label{table_d4su2_stringm}
The lightest mass $am_l$ of a fundamental string wrapped 
around a spatial torus. The first column comes from good
fits to $p=0$ correlators, chosen so as to minimise the 
errors. The second column contains cautious `very safe' 
estimates with larger errors. The third and fourth columns use
both $p=0$ and, where useful, $p\not= 0$ correlators.}
\end{center}
\end{table}

\begin{table}
\begin{center}
\begin{tabular}{|c|c||c|c|c|c|}\hline
\multicolumn{6}{|c|}{ $c_s^{eff}(l,l^{\prime})$ ; SU(2) ; D=3+1} \\ \hline
$L$ & $L^{\prime}$ & $p=0$ & `safe' $p=0$ & low $p$ 
& `safe' low $p$  \\ \hline
 10 & 12 & 0.15(6)  & 0.15(6)  & 0.15(6)  & 0.15(6)  \\
 12 & 14 & 0.79(8)  & 0.69(9)  & 0.81(8)  & 0.69(9)  \\
 14 & 16 & 1.11(12) & 1.26(13) & 1.44(9)  & 1.26(13) \\
 16 & 20 & 1.46(9)  & 1.28(11) & 0.95(8)  & 1.26(10) \\
 20 & 24 & 0.60(15) & 0.92(18) & 0.88(15) & 0.88(15) \\
 24 & 32 & 1.17(32) & 0.66(47) & 1.06(21) & 0.85(24) \\
\hline
\end{tabular}
\caption{\label{table_d4su2_cstring1}
Estimating the effective string correction coefficient from 
the masses of pairs of flux loops of length $l=aL$ and 
$l^{\prime}=a L^{\prime}$
respectively, using eqn(\ref{eqn_polypair}).}
\end{center}
\end{table}

\begin{table}
\begin{center}
\begin{tabular}{|c||cc|cc|cc|}\hline
\multicolumn{7}{|c|}{ $c_s(l\geq l_0)$ ; SU(2) ; D=3+1} \\ \hline
$L_0$ & `safe' $p=0$ & $\chi^2/dof$ & 
low $p$ & $\chi^2/dof$  & `safe' low $p$ & $\chi^2/dof$ \\ \hline
 14 & 1.19(4)  & 1.5 & --       & --  & 1.150(32) & 2.2 \\
 16 & 1.14(6)  & 1.7 & 0.94(4)  & 0.2 & 1.087(52) & 2.2 \\
 20 & 0.88(16) & 0.2 & 0.95(12) & 0.4 & 0.87(11)  & 0.0 \\
\hline
\end{tabular}
\caption{\label{table_d4su2_cstring2}
Estimating the string correction coefficient from a
fit of eqn(\ref{eqn_poly}) to the masses of all the
flux loops  of length $l\geq l_0 = aL_0$. In each case we
show the $\chi^2/dof$ of the best fit.}
\end{center}
\end{table}

\begin{table}
\begin{center}
\begin{tabular}{|c|c||c|c|}\hline
\multicolumn{4}{|c|}{D=3+1} \\ \hline
\multicolumn{2}{|c||}{SU(4)} & \multicolumn{2}{|c|}{SU(5)} \\ \hline
$\beta$ & $a\surd\sigma$ & $\beta$ & $a\surd\sigma$ \\ \hline
10.55 & 0.372 & 16.755 & 0.384 \\
10.70 & 0.306 & 16.975 & 0.303 \\
10.90 & 0.243 & 17.270 & 0.245 \\
11.10 & 0.202 & 17.450 & 0.222 \\
11.30 & 0.170 &  --    &   --  \\
\hline
\end{tabular}
\caption{\label{table_scaled4}
Setting the scale of $a$: the string tension for our
SU(4) and SU(5) lattice calculations in D=3+1.}
\end{center}
\end{table}

\begin{table}
\begin{center}
\begin{tabular}{|c|c|c|c|c|}\hline
\multicolumn{5}{|c|}{D=3+1 ; SU(4)} \\ \hline
$\beta$ & lattice & MC sweeps & $am_{k=1}$ & $am_{k=2}$ \\ 
\hline
10.55 & $8^4$    & $2\times 10^5$ & 0.973(17)  & 1.456(30) \\
10.70 & $6^3 16$ & $5\times 10^4$ & 0.268(8)   & 0.329(12) \\
10.70 & $8^3 12$ & $10^5$         & 0.564(10)  & 0.763(24) \\
10.70 & $10^4$   & $10^5$         & 0.8375(92) & 1.197(18) \\
10.70 & $12^4$   & $10^5$         & 1.033(11)  & 1.456(37) \\
10.70 & $14^4$   & $10^5$         & 1.201(34)  & 1.780(60) \\
10.70 & $16^4$   & $10^5$         & 1.316(78)  & 2.35(27)  \\
10.90 & $12^4$   & $10^5$         & 0.622(7)   & 0.896(11) \\ 
11.10 & $16^4$   & $10^5$         & 0.585(8)   & 0.836(21) \\
11.30 & $20^4$   & $10^5$         & 0.5265(56) & 0.740(16) \\ 
\hline
\end{tabular}
\caption{\label{table_datsu4d4}
Masses of the $k=1$ and $k=2$ flux loops that wind once around the
spatial torus. In D=3+1 SU(4), for the lattices and couplings shown.}
\end{center}
\end{table}

\begin{table}
\begin{center}
\begin{tabular}{|c|c|c|c|c|}\hline
\multicolumn{5}{|c|}{D=3+1 ; SU(5)} \\ \hline
$\beta$ & lattice & MC sweeps & $am_{k=1}$ & $am_{k=2}$ \\ 
\hline
16.755 & $8^4$  & $10^5$            & 1.051(13) & 1.70(7)   \\
16.975 & $10^4$ & $2.0\times 10^5$  & 0.816(12) & 1.239(51) \\ 
17.270 & $12^4$ & $1.4\times 10^5$  & 0.638(9)  & 1.061(28) \\ 
17.450 & $16^4$ & $10^5$            & 0.723(10) & 1.168(62) \\ 
\hline
\end{tabular}
\caption{\label{table_datsu5d4}
Masses of the $k=1$ and $k=2$ flux loops that wind once around the
spatial torus. In D=3+1 SU(5), for the lattices and couplings shown.}
\end{center}
\end{table}

\begin{table}
\begin{center}
\begin{tabular}{|c|c||c|c|}\hline
\multicolumn{4}{|c|}{D=2+1} \\ \hline
\multicolumn{2}{|c||}{SU(4)} & \multicolumn{2}{|c|}{SU(6)} \\ \hline
$\beta$ & $a\surd\sigma$ & $\beta$ & $a\surd\sigma$ \\ \hline
18.0  & 0.442 &  42.0 & 0.436 \\
21.0  & 0.361 &  49.0 & 0.353 \\
28.0  & 0.252 &  60.0 & 0.274 \\
33.0  & 0.208 &  75.0 & 0.211 \\
45.0  & 0.147 & 108.0 & 0.141 \\
60.0  & 0.108 &  --   &  --   \\
\hline
\end{tabular}
\caption{\label{table_scaled3}
Setting the scale of $a$: the string tension for our
SU(4) and SU(6) lattice calculations in D=2+1.}
\end{center}
\end{table}

\begin{table}
\begin{center}
\begin{tabular}{|c|c|c|c|c|}\hline
\multicolumn{5}{|c|}{D=2+1 ; SU(4)} \\ \hline
$\beta$ & lattice & MC sweeps & $am_{k=1}$ & $am_{k=2}$ \\ 
\hline
18.0 & $8^3$     & $2\times 10^5$  & 1.497(18)  & 2.022(66)  \\
21.0 & $10^3$    & $2\times 10^5$  & 1.223(11)  & 1.690(38)  \\
28.0 & $4^2 48$  & $10^5$          & 0.1347(12) & 0.1836(21) \\
28.0 & $6^2 28$  & $10^5$          & 0.2724(15) & 0.3630(26) \\
28.0 & $8^2 16$  & $10^5$          & 0.4276(33) & 0.5652(64) \\
28.0 & $10^2 16$ & $1.5\times10^5$ & 0.5720(30) & 0.7824(63) \\
28.0 & $12^3$    & $2\times 10^5$  & 0.7152(39) & 0.9718(95)  \\
28.0 & $16^3$    & $2\times 10^5$  & 0.9937(58) & 1.345(14)  \\
33.0 & $16^3$    & $2\times 10^5$  & 0.6622(29) & 0.914(6)   \\
45.0 & $24^3$    & $2\times 10^5$  & 0.4974(22) & 0.6815(40) \\
60.0 & $32^3$    & $2\times 10^5$  & 0.3571(14) & 0.4888(15) \\
\hline
\end{tabular}
\caption{\label{table_datsu4d3}
Masses of the $k=1$ and $k=2$ flux loops that wind once around the
spatial torus. In D=2+1 SU(4), for the lattices and couplings shown.}
\end{center}
\end{table}

\begin{table}
\begin{center}
\begin{tabular}{|c|c|c|c|c|c|}\hline
\multicolumn{6}{|c|}{ D=2+1 ; SU(6)} \\ \hline
$\beta$ & lattice & MC sweeps & $am_{k=1}$ & $am_{k=2}$ 
& $am_{k=3}$ \\ \hline
42.0  & $8^3$  & $3\times 10^5$ & 1.453(14)  & 2.51(10)   & 2.85(30)  \\
49.0  & $10^3$ & $2\times 10^5$ & 1.194(9)   & 2.030(42)  & 2.21(9)   \\
60.0  & $12^3$ & $2\times 10^5$ & 0.8575(47) & 1.443(15)  & 1.621(25) \\
75.0  & $16^3$ & $2\times 10^5$ & 0.6825(31) & 1.1305(72) & 1.275(11) \\
108.0 & $24^3$ & $3\times 10^5$ & 0.4552(13) & 0.7534(21) & 0.8424(98) \\
\hline
\end{tabular}
\caption{\label{table_datsu6d3}
Masses of the $k=1$, $k=2$ and $k=3$ flux loops that wind once around 
the spatial torus. In D=2+1 SU(6), for the lattices and couplings shown.}
\end{center}
\end{table}

\begin{table}
\begin{center}
\begin{tabular}{|c||c|c||c|c|}\hline
\multicolumn{5}{|c|}{ D=2+1 ; SU(4) ; $\beta=28$} \\ \hline
\multicolumn{1}{|c||}{} &
\multicolumn{2}{c||}{$am_{k=1}(L)$} & 
\multicolumn{2}{c||}{$am_{k=2}(L)$}  \\ \hline
$L_\perp$ & $L=8$ & $L=12$ & $L=8$ & $L=12$  \\ \hline
2  & 1.188(10) & 1.811(23)  & 1.552(28) & 2.27(10) \\
3  & 0.738(9)  & 1.1900(73) & 1.008(11) & 1.45(15) \\
4  & 0.5093(37) & 0.819(13) & 0.667(9) & 1.04(4) \\
5  & 0.4297(36) & 0.6814(83) & 0.5583(62) & 0.899(12) \\
6  & 0.4207(28) & 0.6965(46) & 0.5504(57) & 0.869(22) \\
8  & 0.4276(33) & 0.710(5)   & 0.5652(64) & 0.947(9)  \\
10 & 0.4274(30) & 0.706(5)   & 0.5770(56) & 0.971(11) \\
12 & 0.4327(30) & 0.715(4)   & 0.5985(58) & 0.972(10) \\
16 & 0.4347(28) & 0.721(5)   & 0.598(7)   & 0.948(24) \\
20 & 0.4331(30) & 0.713(4)   & 0.600(5)   & 0.991(12) \\
\hline
\end{tabular}
\caption{\label{table_widthsu4}
Masses of flux loops that wind once around a spatial torus
of length $L$, as a function of the length, 
$L_\perp$, of the other spatial torus.}
\end{center}
\end{table}

\begin{table}
\begin{center}
\begin{tabular}{|c||c|c||c|c||c|c|}\hline
\multicolumn{7}{|c|}{ D=2+1 ; SU(6) ; $\beta=60$} \\ \hline
\multicolumn{1}{|c||}{} &
\multicolumn{2}{c||}{$am_{k=1}(L)$} & 
\multicolumn{2}{c||}{$am_{k=2}(L)$} & 
\multicolumn{2}{c||}{$am_{k=3}(L)$}  \\ \hline
$l_\perp$ & $L=10$ & $L=12$ & $L=10$ & $L=12$ & $L=10$ & $L=12$ \\ \hline
2  & 1.40(13)  &  2.055(8)  & 2.66(20)  & 2.70(41)   & 3.11(4)    & 2.63(91) \\ 
3  & 1.050(29) &  1.227(56) & 1.785(38) & 2.069(51)  & 1.929(70)  & 2.33(21)  \\ 
4  & 0.692(12) &  0.841(18) & 1.178(14) & 1.419(28)  & 1.360(27)  & 1.611(54) \\ 
5  & 0.6767(66) & 0.8455(65) & 1.064(14) & 1.280(11) & 1.163(16)  & 1.389(31) \\
6  & 0.6915(67) & 0.8462(59) & 1.084(12) & 1.368(23)  & 1.200(21) & 1.531(31) \\
8  & 0.7005(52) & 0.8671(54) & 1.137(14) & 1.390(23)  & 1.194(65) & 1.590(37) \\
10 & 0.7015(49) & 0.843(13)  & 1.180(14) & 1.435(16)  & 1.327(27) & 1.658(55) \\
12 & --         & 0.8575(47) &  --       & 1.443(15)  & --        & 1.621(25) \\
\hline
\end{tabular}
\caption{\label{table_widthsu6}
Masses of flux loops that wind once around a spatial torus
of length $L$, as a function of the length, $L_\perp$, of
the other spatial torus.}
\end{center}
\end{table}

\begin{table}
\begin{center}
\begin{tabular}{|c|c|c|c|c|}\hline
\multicolumn{5}{|c|}{D=3+1 ; SU(4) ; finite T} \\ \hline
$L_t$ & $T/\surd\sigma$ & $\langle l_p \rangle$ 
& $am_t$  & $\langle Q^2 \rangle$  \\ \hline
 2 & 1.63 &  0.5669(1)  & 0.000    & -- \\
 3 & 1.09 &  0.3414(1)  & 0.003    & -- \\ 
 4 & 0.82 &  0.1922(2)  & 0.005    & 0.002(1) \\ 
 5 & 0.65 &  0.0041(40) & 0.052(1) & 0.150(15) \\ 
 6 & 0.54 & -0.0006(10) & 0.230(8) & 1.363(53) \\ \hline
10 & `0.33' & --        & 0.838(9) & 2.56(9)  \\ \hline
\end{tabular}
\caption{\label{table_tempd4a}
Calculations at finite temperature, $T$, on 
$10^3\times L_t$ lattices. In SU(4) at $\beta=10.7$.
We show the thermal line average,  $\langle l_p \rangle$,
the lightest mass coupled to the thermal line, $am_t$,
and the size of the topological fluctuations,
$\langle Q^2 \rangle$.}
\end{center}
\end{table}

\begin{table}
\begin{center}
\begin{tabular}{|c|c|c||c|c|}\hline
\multicolumn{5}{|c|}{D=3+1 ; SU(4) ; finite T} \\ \hline
\multicolumn{3}{|c||}{} &
\multicolumn{2}{|c|}{$\sigma_{k=2}/\sigma$} \\ \hline
$L_t$ & $am_{k=1}$ & $am_{k=2}$ &  $c_s=0$ &  $c_s=1$ \\ \hline

 2 & 2.71(17)  & 2.7(7) & -- & -- \\
 3 & 1.523(28) & 2.11(10) & 1.385(70) & 1.361(66) \\
 4 & 1.044(9)  & 1.40(3)  & 1.341(30) & 1.310(28) \\
 5 & 0.800(9)  & 1.101(20) & 1.376(30) & 1.333(26) \\
 6 & 0.727(16) & 0.90(4)   & 1.238(55) & 1.208(54) \\ \hline
10 & 0.838(9)  & 1.197(18) & 1.429(27) & 1.382(23) \\ \hline
\end{tabular}
\caption{\label{table_tempd4b}
Calculations at the finite temperatures listed in
Table~\ref{table_tempd4a}. We list the `spatial' loop
masses and the corresponding ratio of `spatial' string
tensions, calculated with and without a bosonic
string correction.} 
\end{center}
\end{table}

\begin{table}
\begin{center}
\begin{tabular}{|c|c|c|}\hline
\multicolumn{3}{|c|}{D=2+1 ; SU(4)} \\ \hline
$\beta$ & lattice & $am_{k=2S}$ \\ \hline
28.0 & $12^3$  &  1.601(23)\\
33.0 & $16^3$  &  1.352(15)\\
45.0 & $24^3$  &  0.979(40)\\ 
60.0 & $32^3$  &  0.7694(70)\\ \hline
\end{tabular}
\caption{\label{table_massrepsu4}
Masses of flux loops in the $k=2$ symmetric representation.
In D=2+1 SU(4), for the lattices and couplings shown.} 
\end{center}
\end{table}

\begin{table}
\begin{center}
\begin{tabular}{|c|c|c|c|c|}\hline
\multicolumn{5}{|c|}{D=2+1 ; SU(6)} \\ \hline
$\beta$ & lattice & $am_{k=2S}$ & $am_{k=3M}$ & $am_{k=3S}$ \\ \hline
49.0 & $10^3$  &  2.58(19) & -- & --\\
60.0 & $12^3$  &  1.928(41) & 2.44(12) & --\\
75.0 & $16^3$  &  1.512(15) & 1.845(37) & 2.270(87)\\
108.0 & $24^3$  & 1.0210(47) & 1.274(25) & 1.646(17) \\ \hline
\end{tabular}
\caption{\label{table_massrepsu6}
Masses of flux loops in the $k=2$ symmetric ($k=2S$), $k=3$ mixed ($k=3M$)
and $k=3$ symmetric ($k=3S$) representations.
In D=2+1 SU(6), for the lattices and couplings shown.} 
\end{center}
\end{table}

\clearpage

\begin	{figure}[p]
\begin	{center}
\leavevmode
\epsfig{figure=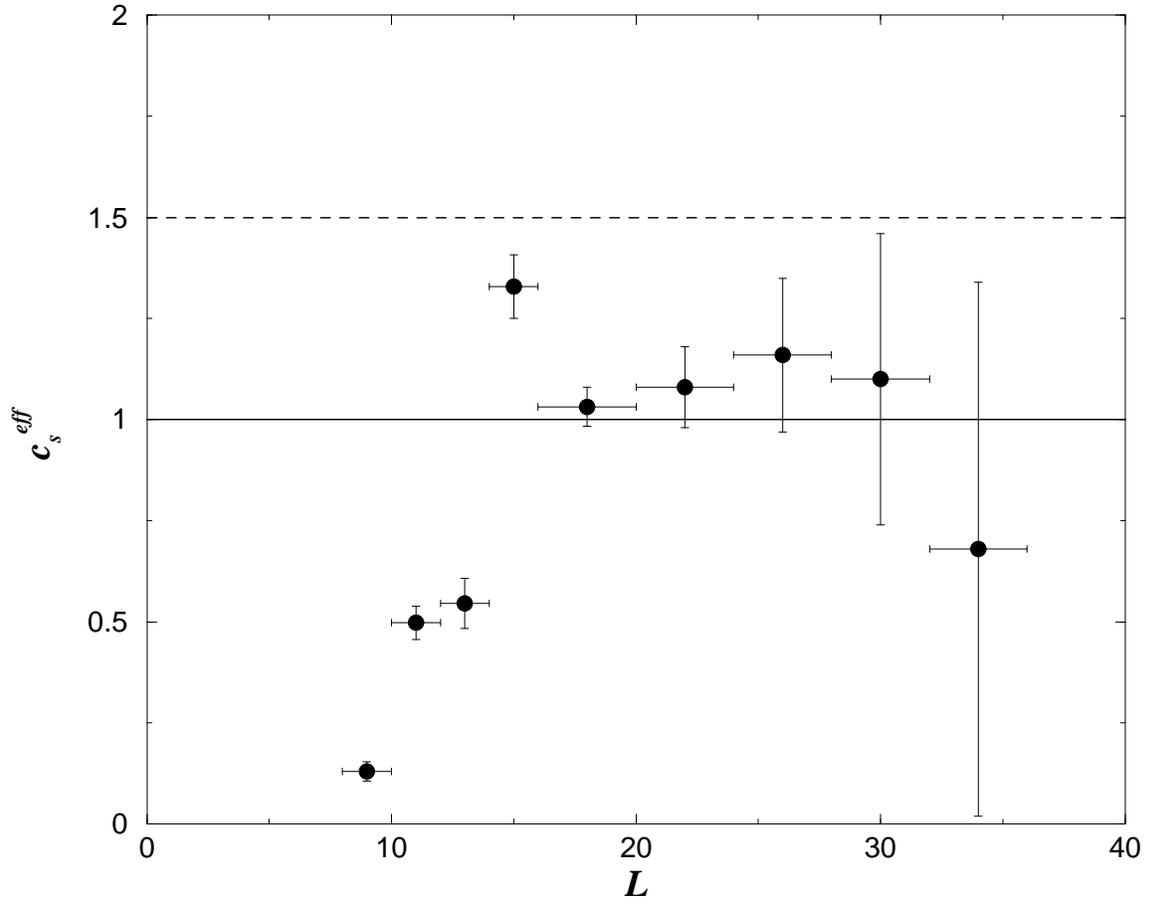, angle=270, width=15cm} 
\end	{center}
\vskip 0.15in
\caption{The  D=2+1 effective string correction coefficient estimated 
from the masses of flux loops of different lengths (indicated
by the span of the horizontal error bar) using 
eqn(\ref{eqn_polypair}). The solid line is what one expects for
a simple bosonic string.}
\label{fig_d3cspair}
\end 	{figure}

\begin	{figure}[p]
\begin	{center}
\leavevmode
\epsfig{figure=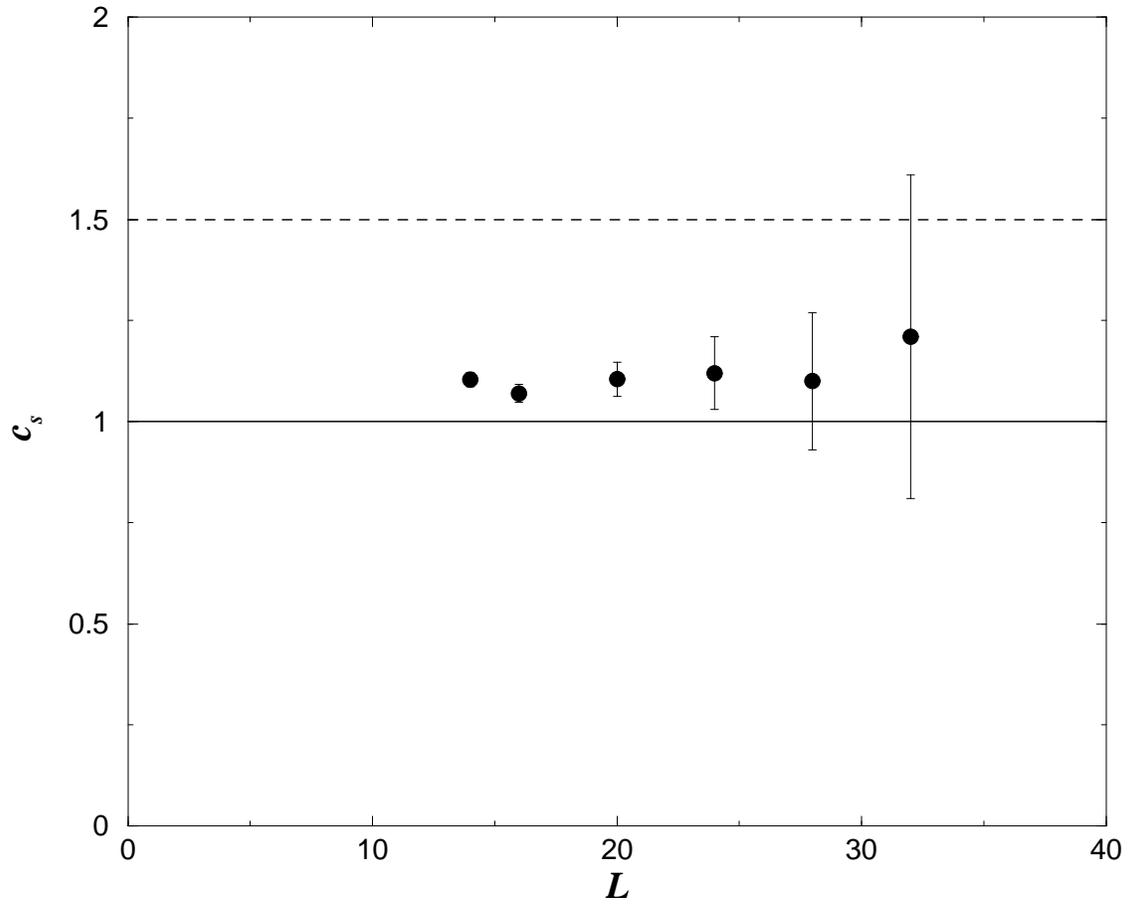, angle=270, width=15cm} 
\end	{center}
\vskip 0.15in
\caption{The  D=2+1 string correction coefficient estimated by
fitting the masses of all flux loops with length greater
than $L$, as a function of $L$.} 
\label{fig_d3csall}
\end 	{figure}

\begin	{figure}[p]
\begin	{center}
\leavevmode
\epsfig{figure=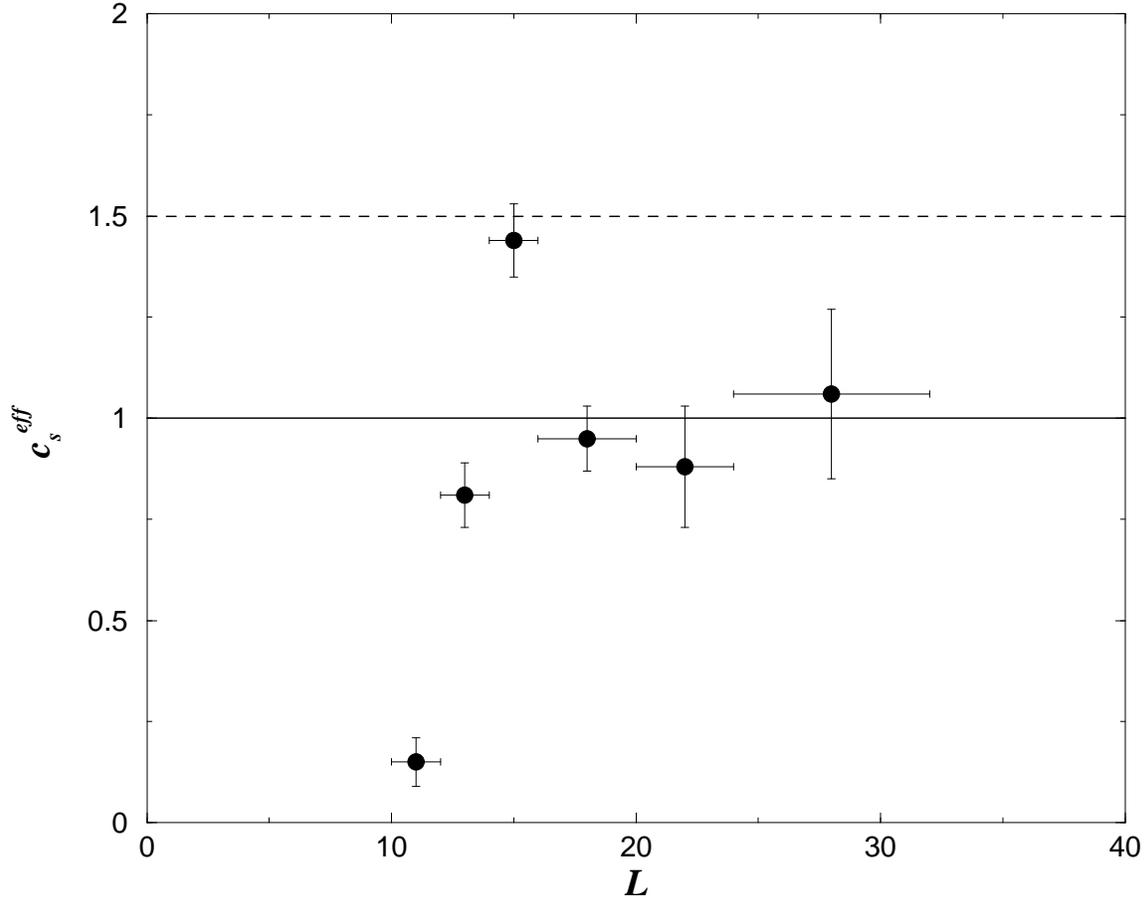, angle=270, width=15cm} 
\end	{center}
\vskip 0.15in
\caption{The D=3+1 effective string correction coefficient estimated 
from the masses of flux loops of different lengths (indicated
by the span of the horizontal error bar) using 
eqn(\ref{eqn_polypair}). The solid line is what one expects for
a simple bosonic string. For comparison the dashed line indicates the
value for the Neveu-Schwartz string. We use masses from the
third column in Table~\ref{table_d4su2_cstring1}.}
\label{fig_d4cspair}
\end 	{figure}

\begin	{figure}[p]
\begin	{center}
\leavevmode
\epsfig{figure=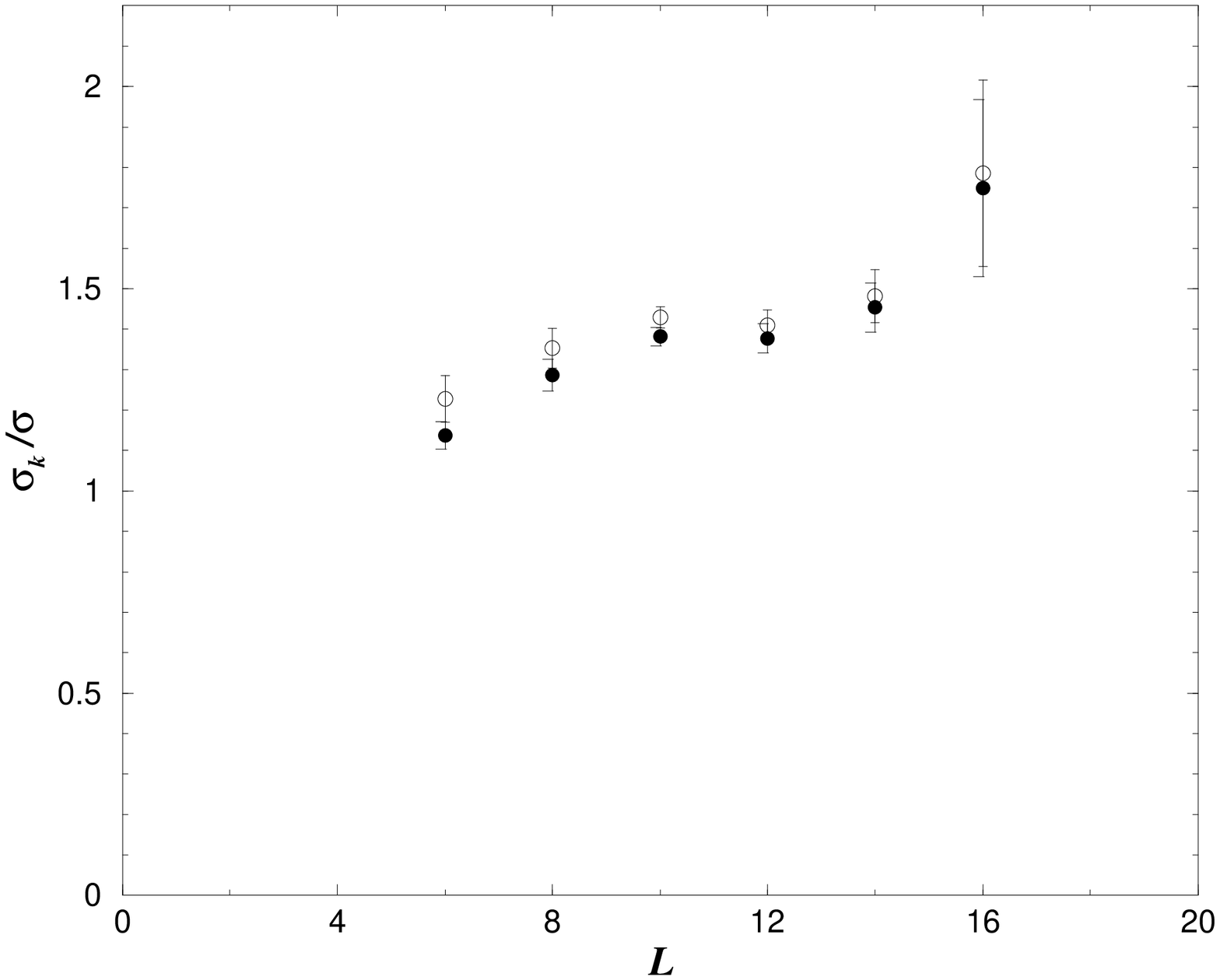, angle=270, width=15cm} 
\end	{center}
\vskip 0.15in
\caption{The ratio of $k=2$ and $k=1$ string tensions in 
D=3+1 SU(4) at $\beta=10.7$ extracted from flux loops of
length $l=aL$. We show values extracted using a bosonic 
string correction, ($\bullet$), and no string correction 
at all ($\circ$).} 
\label{fig_d4su4string}
\end 	{figure}

\begin	{figure}[p]
\begin	{center}
\leavevmode
\epsfig{figure=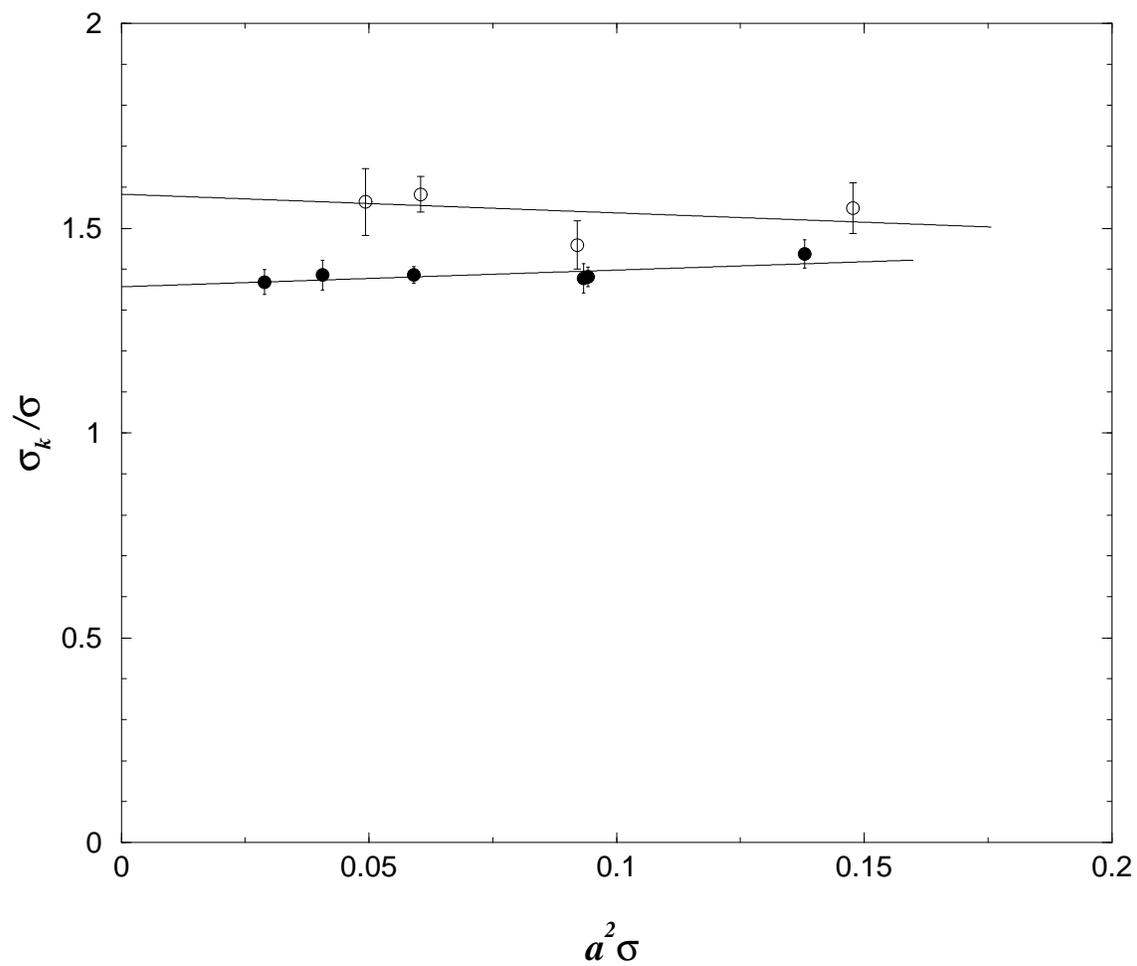, angle=270, width=15cm} 
\end	{center}
\vskip 0.15in
\caption{The ratio of $k=2$ and $k=1$ string tensions in 
our D=3+1 SU(4) ($\bullet$) and SU(5) ($\circ$) lattice
calculations plotted as a function of $a^2\sigma$.
Extrapolations to the continuum limit, using a leading 
$O(a^2)$ correction, are displayed.}
\label{fig_d4sig2}
\end 	{figure}

\begin	{figure}[p]
\begin	{center}
\leavevmode
\epsfig{figure=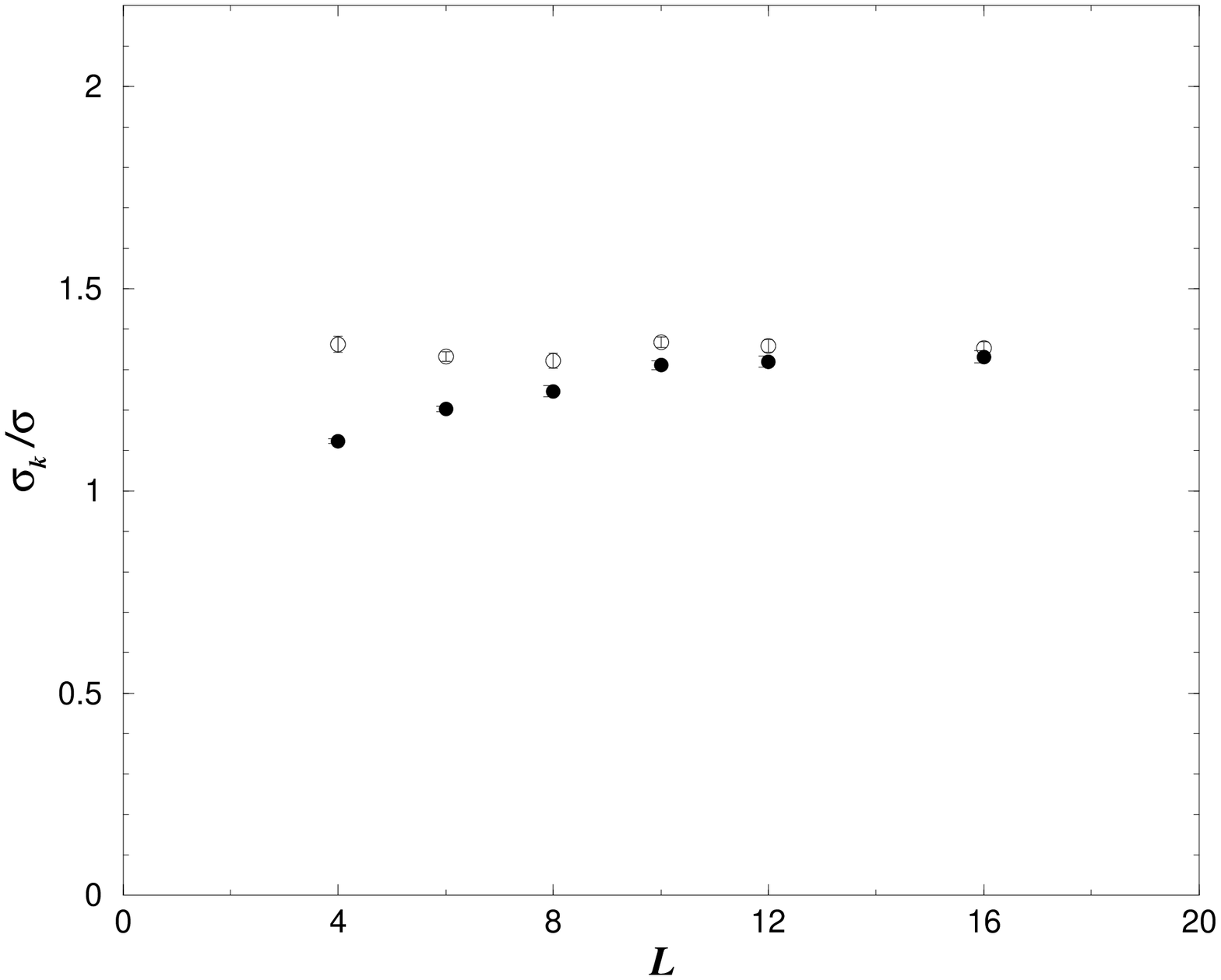, angle=270, width=15cm} 
\end	{center}
\vskip 0.15in
\caption{The ratio of $k=2$ and $k=1$ string tensions in 
D=2+1 SU(4) at $\beta=28.0$ extracted from flux loops of
length $l=aL$. We show values extracted using a bosonic 
string correction, ($\bullet$), and no string correction 
at all ($\circ$).} 
\label{fig_d3su4string}
\end 	{figure}

\begin	{figure}[p]
\begin	{center}
\leavevmode
\epsfig{figure=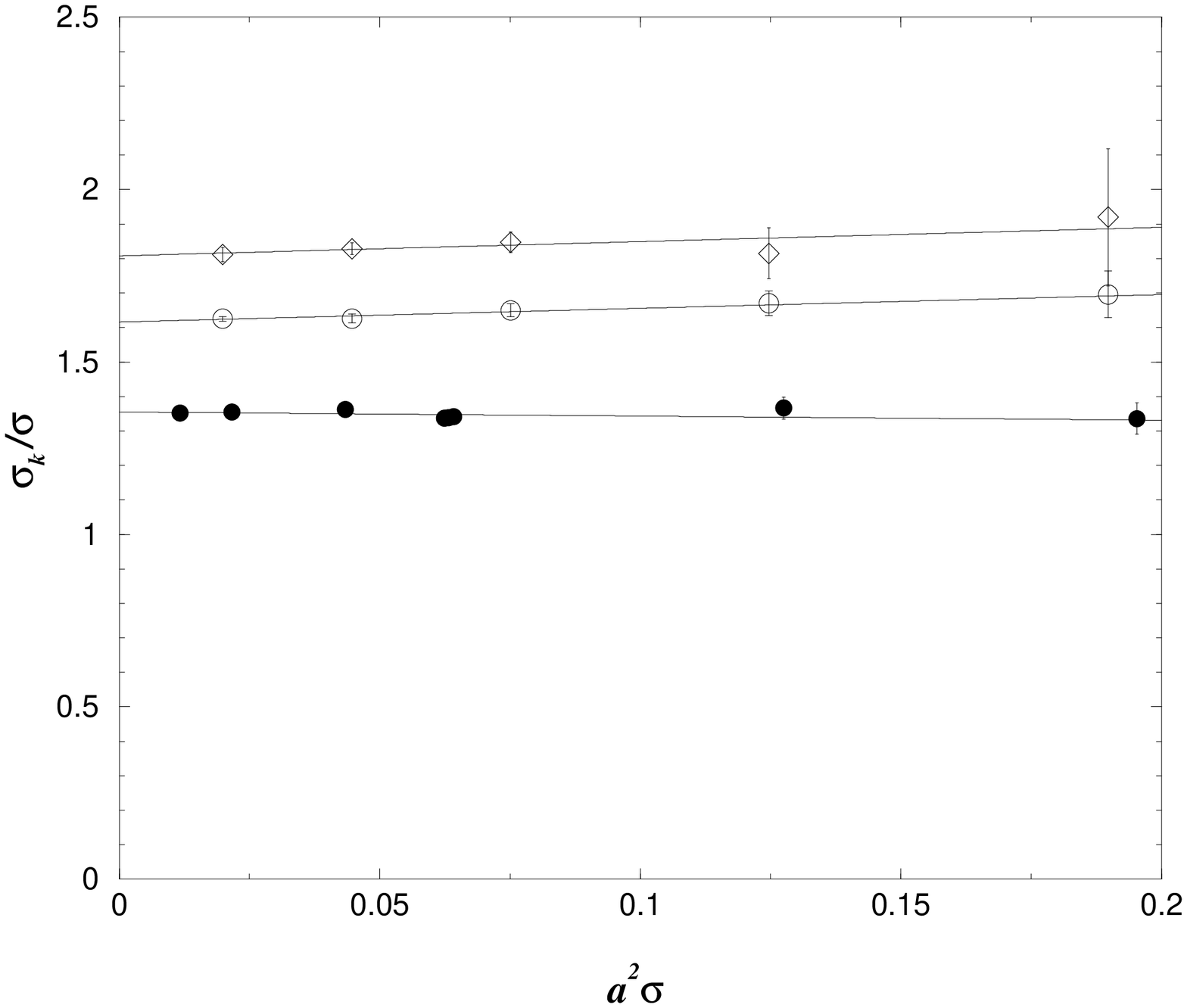, angle=270, width=15cm} 
\end	{center}
\vskip 0.15in
\caption{The ratio of $k=2$ and $k=1$ string tensions in 
our D=2+1 SU(4) ($\bullet$) and SU(6) ($\circ$) lattice
calculations plotted as a function of $a^2\sigma$.
Also shown is the $k=3$ to $k=1$ ratio ($\diamond$) in SU(6).
Extrapolations to the continuum limit, using a leading 
$O(a^2)$ correction, are displayed.}
\label{fig_d3sigk}
\end 	{figure}

\begin	{figure}[p]
\begin	{center}
\leavevmode
\epsfig{figure=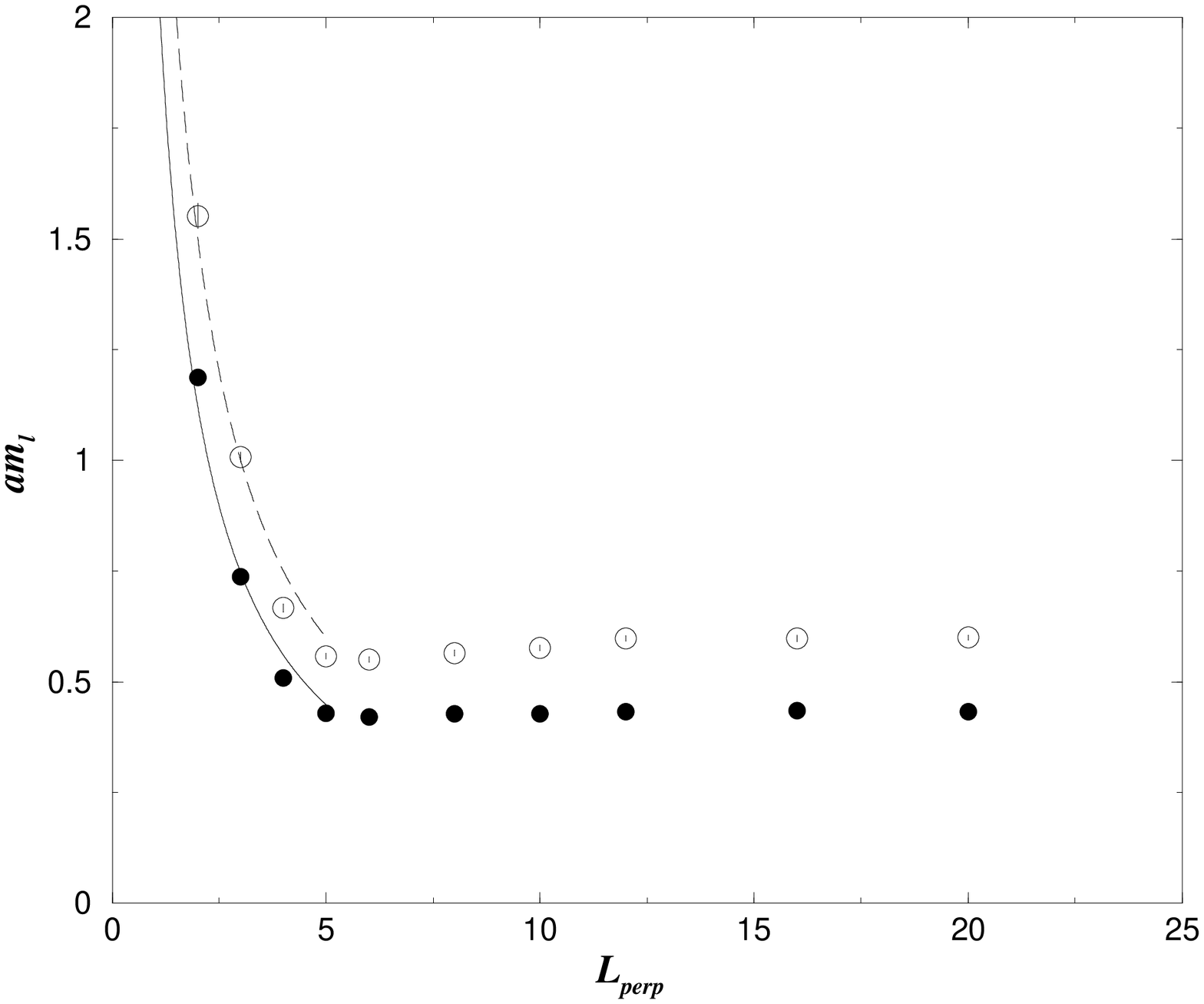, angle=270, width=15cm} 
\end	{center}
\vskip 0.15in
\caption{The masses of the $k=1$ and $k=2$ flux loops
of length $L=8$ in the D=2+1 SU(4) gauge theory at 
$\beta=28$, versus the size of the transverse spatial torus
$L_{perp} \equiv L_{\perp}$. Shown is the dependence in
eqn(\ref{eqn_mwidth}) fitted to the smallest values of $L_{perp}$.}
\label{fig_widthsu4d3}
\end 	{figure}

\begin	{figure}[p]
\begin	{center}
\leavevmode
\epsfig{figure=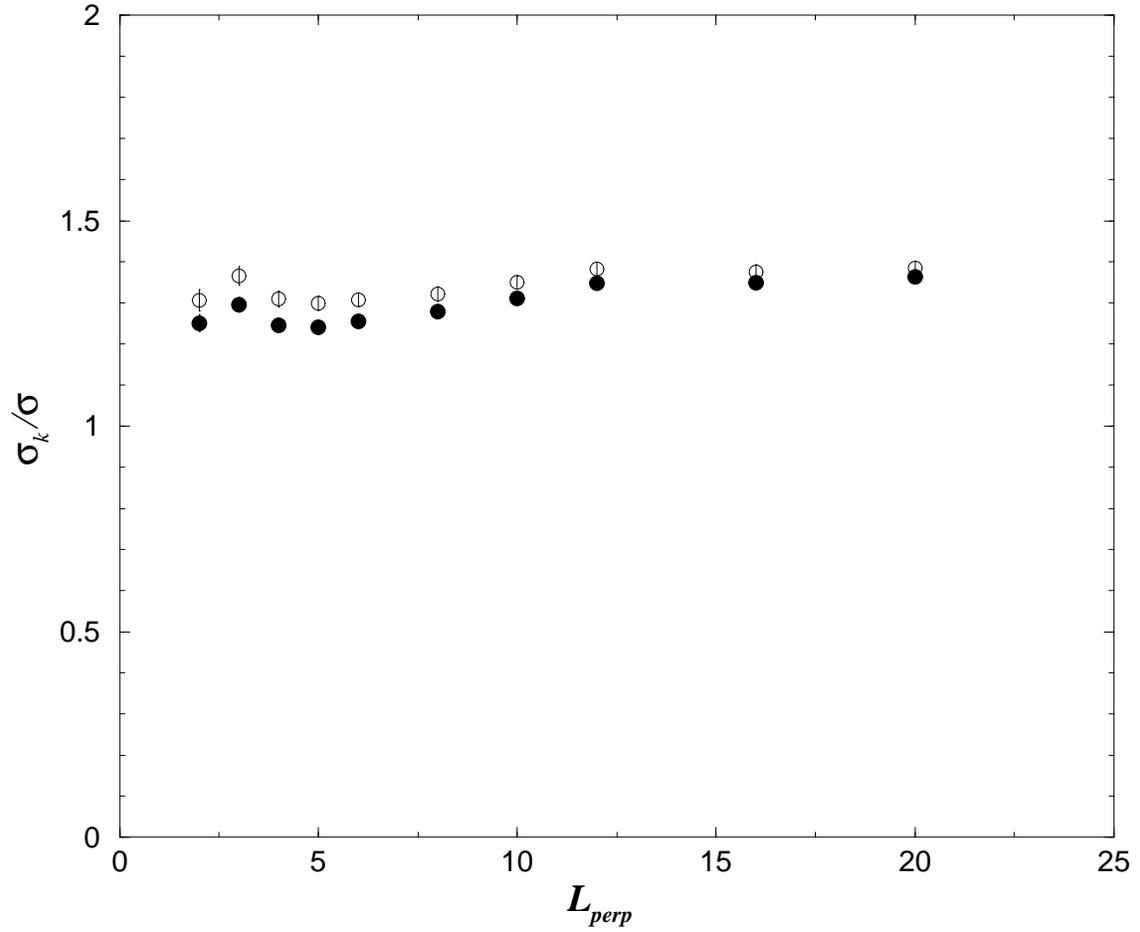, angle=270, width=15cm} 
\end	{center}
\vskip 0.15in
\caption{The ratio of the $k=2$ to $k=1$ string tensions
in the D=2+1 SU(4) gauge theory at $\beta=28$ versus the size 
of the transverse spatial torus $L_{perp} \equiv L_{\perp}$. With
($\bullet$) and without ($\circ$) a (bosonic) string
correction.}
\label{fig_rwidthsu4d3}
\end 	{figure}

\begin	{figure}[p]
\begin	{center}
\leavevmode
\epsfig{figure=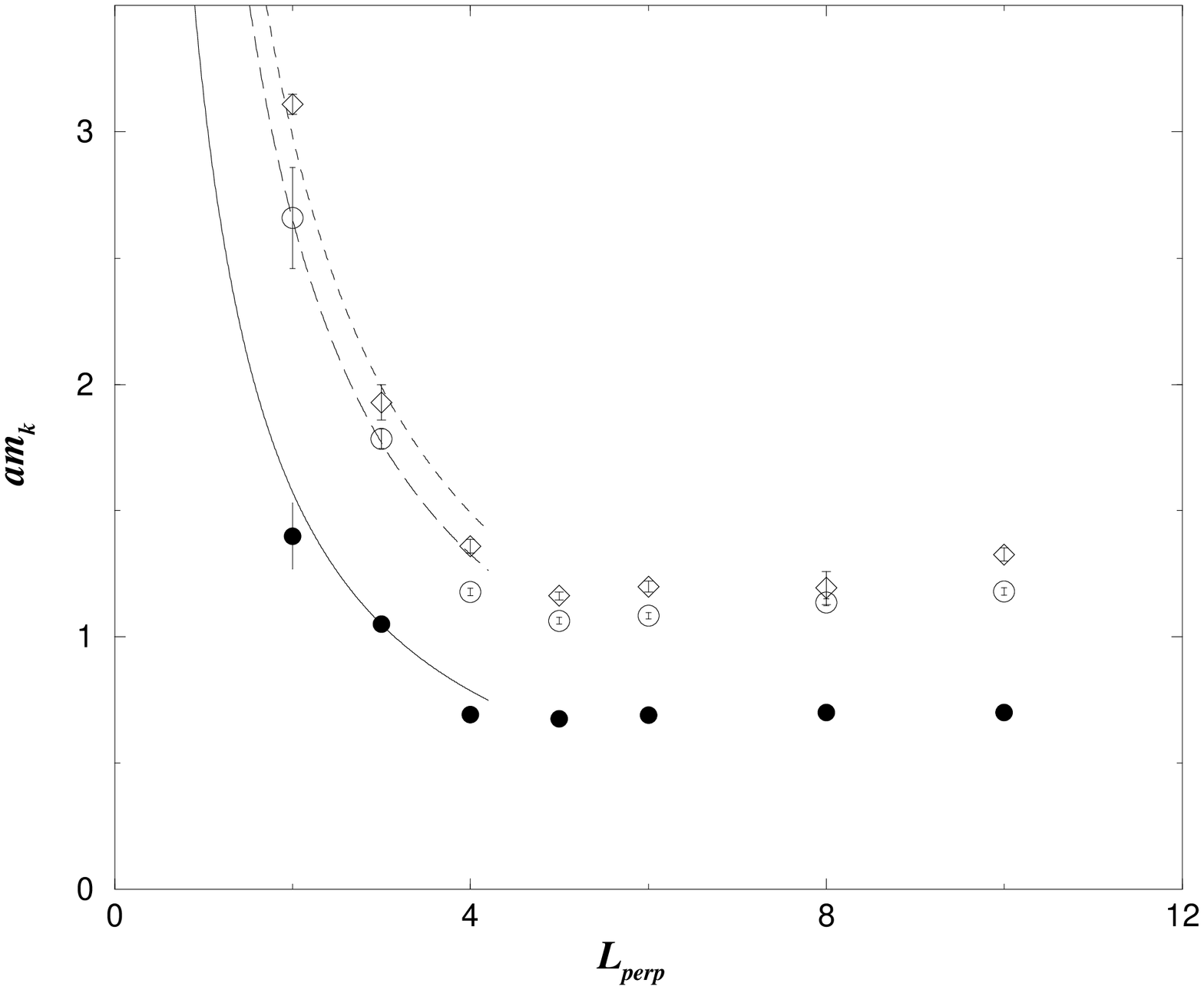, angle=270, width=15cm} 
\end	{center}
\vskip 0.15in
\caption{The masses of the $k=1$, $k=2$ and $k=3$ flux loops
of length $L=10$ in the D=2+1 SU(6) gauge theory at 
$\beta=60$ versus the size of the transverse spatial torus
$L_{perp} \equiv L_{\perp}$. Shown is the dependence in 
eqn(\ref{eqn_mwidth}) fitted to the smallest values of $L_{perp}$.}
\label{fig_widthsu6d3}
\end 	{figure}

\begin	{figure}[p]
\begin	{center}
\leavevmode
\epsfig{figure=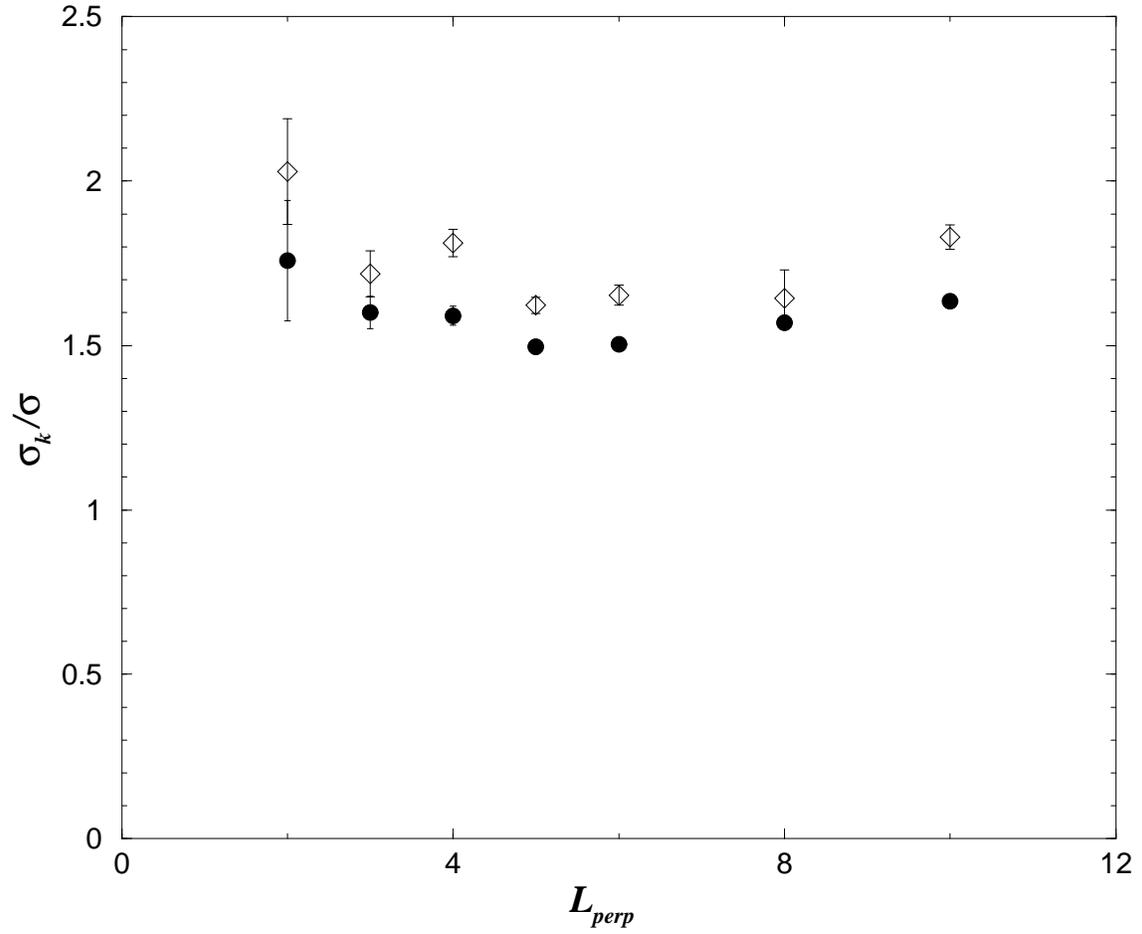, angle=270, width=15cm} 
\end	{center}
\vskip 0.15in
\caption{The ratio of the $k=2$ to $k=1$ ($\bullet$)
and $k=3$ to $k=1$ ($\diamond$) string tensions
in the D=2+1 SU(6) gauge theory at $\beta=60$ versus the 
size of the transverse spatial torus $L_{perp} \equiv L_{\perp}$. 
The (bosonic) string correction has been included.}
\label{fig_rwidthsu6d3}
\end 	{figure}

\end{document}